\documentclass[paper]{ptephy}

\usepackage{boites}
\usepackage{graphicx}
\usepackage{bm}
\usepackage{amsmath,amssymb,amsfonts}

\newcommand{\nc}{\newcommand}		
\nc{\vc}[1]	{\mbox{\boldmath $#1$}}	
\nc{\del}       {\partial}              
\nc{\bra}       {\langle}               
\nc{\ket}       {\rangle}               
\nc{\bras}[1]   {\langle #1|}           
\nc{\kets}[1]   {|#1\rangle}            
\nc{\mapleft}[1]{			
 \smash{\mathop{\,			%
  \hbox to 1.5cm{\rightarrowfill}\, }\limits_{#1}}}
\nc{\beq}     {\begin{eqnarray}}
\nc{\eeq}    {\end{eqnarray}}
\nc{\nn}   {\\\nonumber}
\nc{\dl} {\delta}
\nc{\tht} {\theta}
\nc{\sig} {\sigma}
\nc{\Dl} {\Delta}
\nc{\Sig} {\Sigma}
\nc{\dg}  {\dagger}
\nc{\ti} {\tilde}
\nc{\lm} {\lambda}
\nc{\bx}     {\bold x}
\nc{\br} {\bold r}
\nc{\fra}     {\frac{1}{2}}
\nc{\AMD}{{\rm AMD}}

\begin{document}

\title{Tensor-optimized antisymmetrized molecular dynamics in nuclear physics}

\author{\name{Takayuki Myo}{1,2}, \name{Hiroshi Toki}{2},  \name{Kiyomi Ikeda}{3}, \name{Hisashi Horiuchi}{2}, and \name{Tadahiro Suhara}{4}}

\address{\affil{1}{General Education, Faculty of Engineering, Osaka Institute of Technology, Osaka, Osaka 535-8585, Japan}
\affil{2}{Research Center for Nuclear Physics (RCNP), Osaka University, Ibaraki, Osaka 567-0047, Japan}
\affil{3}{RIKEN Nishina Center, Wako, Saitama 351-0198, Japan}
\affil{4}{Matsue College of Technology, Matsue 690-8518, Japan}
\email{myo@ge.oit.ac.jp~~~takayuki.myo@oit.ac.jp}}

\begin{abstract}%
We develop a new formalism to treat nuclear many-body systems using bare nucleon-nucleon interaction.  It has become evident that the tensor interaction plays important role in nuclear many-body systems due to the role of the pion in strongly interacting system. We take the antisymmetrized molecular dynamics (AMD) as a basic framework and add a tensor correlation operator acting on the AMD wave function using the concept of the tensor-optimized shell model (TOSM).  We demonstrate a systematical and straightforward formulation utilizing the Gaussian integration and differentiation method and the antisymmetrization technique to calculate all the matrix elements of the many-body Hamiltonian.  We can include the three-body interaction naturally and calculate the matrix elements systematically in the progressive order of the tensor correlation operator.  We call the new formalism ``tensor-optimized antisymmetrized molecular dynamics''. 
\end{abstract}

\subjectindex{D10, D11}


\maketitle

\section{Introduction}\label{sec:intro}
The tensor interaction plays a very important role in nuclear physics. It originates from the pion exchange between nucleons, which is the most essential component in the nucleon-nucleon interaction.  The Green's function Monte Carlo (GFMC) simulation by the Argonne group for light nuclei demonstrated that the pion contribution on the nuclear binding energy is about 80\% of the entire contribution of the two-body interaction~\cite{pieper01}.  The pion exchange interaction can be divided into tensor and central spin-spin components.  If we take a phenomenological potential as the Argonne AV8$^\prime$ potential~\cite{wiringa84, wiringa95}, we see that the lightest nucleus, a deuteron, cannot be bound by the central interaction alone: the tensor interaction plays a dominant role in the binding of the deuteron~\cite{ikeda10}.  Analyzing the energy contribution of the tensor interaction, we find that the transition matrix element between the s-wave and d-wave components provides the largest attraction for the deuteron binding.

This finding motivated the introduction of the tensor-optimized shell model (TOSM), where the TOSM wave function has a low-momentum shell-model state and also high-momentum 2-particle-2-hole (2p-2h) configurations, aiming at describing heavier nuclei and nuclear matter~\cite{myo09}.  The 2p-2h configurations are excited by the tensor interaction from the low-momentum shell-model state, which induces high-momentum transfer between nucleons, and provides a large attraction energy.  The TOSM was applied to light mass nuclei as He, Li and Be isotopes~\cite{myo11, myo12, myo14}.  The level orders and spacings come out to be quite good due to the adequate role of the tensor interaction, although the absolute values for the binding energies are not reproduced due to the lack of the three-body interaction~\cite{myo11, myo12, myo14}.  The saturation property seems satisfied for shell-model states with the use of the bare nucleon-nucleon interaction.  The TOSM was able to describe shell-model states and also generate low-lying alpha cluster structures in the spectrum of $^8$Be~\cite{myo14}.  However, the TOSM could not reproduce sufficient of alpha correlation, and we ought to improve the TOSM for multi-cluster states.   The concept of TOSM was also applied to the few-body framework and demonstrated its goodness~\cite{horii12}.  However, the few-body framework that uses the relative coordinates has  difficulty in handling the antisymmetrization for p-shell nuclei.

On the other hand, antisymmetrized molecular dynamics (AMD) was developed by the Kyoto group with great success to describe both the cluster and shell structures simultaneously~\cite{kanada95, kanada03, kanada12}.  Here, the nuclear dynamics was controlled by an effective interaction, which is obtained from various experimental data.  There are several cases where the effective interaction has to be changed for various observables of nuclear structure.  A typical case for this change is the ground-state energies of $^{12}$C and $^{16}$O.  It is highly desirable to use the bare nucleon-nucleon interaction in the AMD framework for the description of light nuclei, where the structural change between shell and cluster states is essential.  For $^8$Be, both the shell and alpha structures are observed experimentally and the AMD description should be essential for good description of this nucleus~\cite{myo14}.

It has become clear from a few-body study and the GFMC simulations for light nuclei that the three-body interaction has to be introduced for a satisfactory description of finite-mass nuclei~\cite{pieper01}.  The three-body interaction is mainly caused by the pion exchange interaction leading to delta-isobar excitation~\cite{fujita57}.  Hence, its structure is again dominantly described by the successive tensor interaction.  Such a three-body interaction usually makes calculations of matrix elements highly complicated.  With the success of the TOSM for light nuclei and  the ability of the AMD formulation, it is a good idea to combine these two merits for the description of nuclear structure using the bare nucleon-nucleon interaction.  After many trials, we arrived at constructing a powerful method to treat nuclear many-body systems based on the AMD wave function and added a tensor component by applying the tensor correlation operator on the AMD wave function.  We treat the tensor correlation operator as an operator on two nucleons of many nucleon systems.  If we want to take the matrix elements of the two-body interaction of the tensor-correlated AMD wave function, it turns out to calculate up to 6-body operators.  We do these calculations systematically by utilizing the Gaussian integrals, which could be done analytically.  Hence, we are able to handle any number of multi-body operators systematically.  The genuine three-body interaction of the Argonne group is naturally described in the new formalism in the same way as tensor-correlated two-body operators.  We call ``this framework tensor-optimized antisymmetrized molecular dynamics'' (TOAMD).

There are three essential technologies in TOAMD to calculate multi-body operators based on the Gaussian wave functions.  The first important step is to write all the interactions and correlations in addition to the wave functions as a sum of Gaussian functions.  We are then able to take all the necessary integrations of the Gaussian functions analytically.  The second important step is to take care of the antisymmetrization.  For this difficult problem, we write all the interactions and correlations in the momentum space so that we are able to write all operators in separable forms in the particle coordinates~\cite{goto79}.  We are then able to use the matrix technique to treat the antisymmetrization systematically.  The third important ingredient is to treat all the necessary momentum integrations with a multiple of momenta in the Gaussian integrations.  We introduce the source terms for the momenta and take derivatives of the integrated result of the fundamental Gaussian integrals.  All calculations of the matrix elements are performed in the rectangular coordinates of particles using the above important ingredients for multi-body operators.  

In light nuclei there appear cluster structures, which are difficult to describe in the shell $+$ mean-field approximation.  The cluster structure is better described in the AMD framework using effective interactions, where the $^4$He wave function is written as $(0s)^4$ Gaussian functions.  Although some trial was performed to include higher-spin states in the AMD framework for $^4$He, it was not successful in treating the tensor interaction~\cite{dote06}.  On the other hand, the TOSM was able to treat the tensor interaction in the shell-model basis by introducing high-momentum 2p-2h configurations.  Use of the bare nucleon-nucleon interaction including the strong tensor interaction in the AMD framework is highly anticipated. This is achieved in the present TOAMD formulation, which is systematical and straightforward to apply to heavier-mass nuclei than those of the GFMC simulation.  We would like to apply the TOAMD for the description of the coexistence of and competition between shell and cluster structures.  One important application of the TOAMD is the alpha-condensed states in $^8$Be, $^{12}$C, and $^{16}$O using the bare nucleon-nucleon interaction~\cite{tohsaki01}.

This paper is arranged as follows.  In Sect.\,\ref{sec:model}, we introduce the tensor-optimized antisymmetrized molecular dynamics (TOAMD), where the TOAMD wave function and the Hamiltonian are written explicitly for nuclear many-body systems.  Here, we write all the matrix elements of the Hamiltonian in the AMD wave function so that we define all ingredients of the TOAMD theory.  In Sect.\,\ref{sec:one_fd}, we calculate matrix elements of an interaction with one tensor correlation operator, which leads to multi-body operators up to four-body for the AMD wave function.  In Sect.\,\ref{sec:momentum}, we write all the necessary integrals and differential formulas for multiple-momentum integrations of Gaussian functions.  In Sect.\,\ref{sec:two_fd}, we calculate two-body interactions with two tensor correlation operators as examples of using all the formula developed for the TOAMD theory.  We describe a systematic method to write matrix elements for many-body systems. In Sect.\,\ref{sec:short}, we introduce the short-range correlation operator as the sum of Gaussian functions. We explicitly give some of the matrix elements for a three-body interaction with the short range and tensor correlations. We present here a systematic method to calculate any complicated matrix elements in the TOAMD theory.  In Sect.\,\ref{sec:summary}, we summarize the present paper. We further present two appendices. In Appendix \ref{sec:matrix}, we desribe the co-factor matrix theory to treat the antisymmetrization.  In Appendix \ref{sec:gauss}, we explicitly give the Gaussian integrals for multi-body operators.

\section{Tensor-optimized antisymmetrized molecular dynamics}\label{sec:model}
We describe here the construction of the tensor-optimized antisymmetrized molecular dynamics (TOAMD).  In this section, we give all the ingredients such as the TOAMD wave function, Hamiltonian, and the matrix elements of the Hamiltonian in the AMD wave function.

\subsection{Wave function of TOAMD}
In this subsection, we introduce the TOAMD wave function for finite nuclei.  In the concept of the tensor-optimized shell model (TOSM), it is important to prepare a basic wave function to represent a correct density profile with low-momentum components, and add high-momentum components to be excited by the strong tensor interaction~\cite{myo09}.  We take the following form for the TOAMD wave function:
\beq
\label{toamdwf}
\kets{\Psi} = \kets{\AMD} + F_D \kets{\AMD}~.
\eeq
Here, $|\AMD\ket$ is an AMD wave function for mass number $A$:
\beq
|\AMD\ket= {\mathcal A}\left\{ \prod_{i=1}^A\psi_{p_i}(\vec r_i)\chi_{p_i}(s_i)\xi_{p_i}(t_i) \right\}
= \frac{1}{\sqrt{A!}} \kets{\det |p_1 p_2 \cdots p_A|}~,
\eeq
where $p_i$ denotes various quantum numbers of a single nucleon and $|p_i \ket= |\psi_{p_i}\chi_{p_i}\xi_{p_i}\ket$. The antisymmetrizer $\mathcal A$ makes sure that the exchange of particle coordinates among all particles have an opposite sign from the original wave function (Slater determinant).  We can include multiple tensor correlation operators $F_D\cdots F_D|\AMD\ket$ for the description of the multi-cluster states and short-range correlation to be discussed later.  The spatial wave function of single nucleon $\psi_{p_i}(\vec{r}_i)$ is written in terms of the shifted Gaussian function:
\beq
\psi_{p_i}(\vec r_i)=\left(\frac{2\nu}{\pi}\right)^{3/4} e^{-\nu(\vec r_i-\vec D_{p_i})^2}~.
\eeq
The particle coordinate is written as $\vec r_i$ for all nucleons $i=1,\ldots,A$.  The size parameter $\nu$ and the position parameter $\vec D_{p_i}$ are the variational parameters to specify the spatial wave function.  The position vectors $\vec D_{p_i}$ are in general complex variables, but we write them as real variables in this paper to simplify the notation.  For numerical calculations it is important to take the AMD wave function in order to calculate matrix elements analytically using the Gaussian integral formula.  As the nuclear system becomes heavy, we ought to include more Gaussian functions and additionally perform angular momentum and parity projections.

The spin wave function $\chi_{p_i}(s_i)$ is written as:
\beq
\chi_{p_i}(s_i)=\beta_{p_i}|\uparrow\ket +(1-\beta_{p_i})|\downarrow\ket~.
\eeq
The spin wave function $\chi_p(s)$ is written as a linear combination of spin-up $|\uparrow\ket$ and spin-down $|\downarrow\ket$ wave functions, where $\beta_p$ is a complex variational parameter. 
As for the isospin part $\xi_{p_i}(t_i)$, we take pure proton and neutron states:
\beq
\xi_{p_i}(t_i)=|{\rm proton}\ket ~~\mbox{or}~~ |{\rm neutron}\ket~.
\eeq
Here, $| {\rm proton}\ket$ and $| {\rm neutron}\ket$ are pure proton and neutron states, respectively.

The tensor correlation operator is expressed as:
\beq
\label{cten}
F_D=\fra \sum_{i\ne j}F_{ij}=\fra \sum_{i\ne j}f_D(r_{ij})S_{12}(r_{ij})\tau_i\cdot \tau_j~,
\eeq
with the variational function written by the sum of Gaussian functions:
\beq
f_D(r_{ij})=\sum_\mu C_\mu r_{ij}^2 e^{-a_\mu r_{ij}^2}~,
\eeq
where $C_\mu$ and $a_\mu$ are variational parameters.
Here the tensor operator is written as:
\beq
S_{12}(r_{ij})=3(\sigma_i\cdot \hat r_{ij})(\sigma_j \cdot \hat r_{ij})-\sigma_i \cdot \sigma_j~.
\eeq
The spin operators $\sigma_i$ and $\sigma_j$ are for particles $i$ and $j$.  The isospin $\tau_i\cdot \tau_j$ operator is added for the tensor correlation operator, where the tensor correlation is very strong because its origin is one-pion exchange.  Although we do not show it explicitly, we also consider the isospin-independent tensor correlation in the calculation.  The relative coordinate $\vec r_{ij}$ is the difference between the positions $\vec r_i$ and $\vec r_j$ of two particles $i$ and $j$, and $\hat r_{ij}$ is a unit vector with its direction.  It is essential to include the tensor-correlated wave function $F_D|\AMD\ket$ explicitly so that the strong tensor interaction, exciting $|\AMD\ket$ to the $F_D|\AMD\ket$ state, provides a large attractive contribution to the total energy.  All the parameters in the wave function (\ref{toamdwf}) are variational parameters.  They are fixed by the energy minimization of the many-body Hamiltonian:
\beq
E=\frac{\bra \AMD|(1+F_D)H(1+F_D)|\AMD \ket}{\bra \AMD|(1+F_D)(1+F_D)|\AMD \ket}~.
\eeq
We have to calculate all the necessary matrix elements for the two and three-body interactions using the TOAMD wave function.  This form of the wave function~(\ref{toamdwf}) was studied for the $^4$He nucleus by Nagata et al. to study the role of the tensor interaction~\cite{nagata59}.

The overlap integral of the AMD wave function is written as:
\beq
\label{det}
\bra \AMD|\AMD \ket
&=&\bra p_1 p_2 p_3 \cdots | \det |q_1 q_2 q_3\cdots|\ket \nn&=&|B|
~=~\left | \begin{array}{ccccc}
\bra p_1|q_1 \ket & \bra p_1|q_2 \ket & \cdots & \bra p_1|q_A\ket\\
\bra p_2|q_1 \ket & \bra p_2|q_2 \ket & \cdots & \bra p_2|q_A\ket\\
\vdots            & \vdots            & \ddots & \vdots \\
\bra p_A|q_1 \ket & \bra p_A|q_2 \ket & \cdots & \bra p_A|q_A\ket\\
\end{array}\right |
\eeq
The single-particle matrix elements are written as:
\beq
\bra p| q \ket=\bra \psi_p| \psi_q \ket\bra \chi_p| \chi_q \ket\bra \xi_p| \xi_q \ket~.
\eeq
The spatial matrix element is:
\beq
\bra \psi_p| \psi_q \ket=e^{-\fra \nu(\vec D_p-\vec D_q)^2}~.
\eeq
As for the spin part, we use the following notation for the matrix element:
\beq
\label{spinm}
M^{pq}=\bra \chi_p|\chi_q \ket=\beta_p^*\beta_q+(1-\beta_p^*)(1-\beta_q)~.
\eeq
For the isospin part, we use the following notation:
\beq
\label{isospinm}
&&\bar M^{pq}=\bra \xi_p|\xi_q \ket=1~~\mbox{or}~~0~,
\eeq
where the states $p$ and $q$ are both the proton states or neutron states for $\bar M^{pq}=1$, and they are different states for $\bar M^{pq}=0$.  Altogether the overlap matrix element is written as:
\beq
\bra p| q \ket=e^{-\fra \nu(\vec D_p-\vec D_q)^2}M^{pq}\bar M^{pq}~.
\eeq
We will give the overlap matrix element of the tensor-correlated wave function $F_D|\AMD\ket$ later, where we describe the method of handling multiple tensor correlation operators.  In the next subsection, we introduce the Hamiltonian and later calculate the matrix element of the Hamiltonian for the AMD wave function $\bra \AMD|H| \AMD \ket$.

\subsection{Nucleon-nucleon interaction and three-body interaction}
We take the many-body Hamiltonian as a summation of the kinetic, two-body and three-body interactions:
\beq
H=T+V+U~.
\eeq
Here, $T$ is the many-body kinetic energy:
\beq
T=\sum_{i=1}^A \left(-\frac{\vec \nabla_i^2}{2m}\right)-T_{\rm c.m.}~.
\eeq 
Here, we use natural units: $\hbar=c=1$.
The first term is the sum of the individual kinetic energies and $T_{\rm c.m.}$ is the center-of-mass (c.m.) kinetic energy.  Hence, $T$ denotes the kinetic energy of the intrinsic motion of nucleons.

Wiringa {\it et al.} constructed phenomenological two-body nucleon-nucleon ($NN$) interactions without and with $\Dl$(1232) degrees of freedom, the Argonne v14 (AV14) and v28 (AV28) models, respectively~\cite{wiringa84, wiringa95}.  In principle we are able to treat any interactions, but here we write all the necessary ingredients to treat the AV14 potential.  We briefly review the content of the $NN$ interactions of the AV14 potential. We consider the $\Dl$(1232) degrees of freedom in terms of a three-body interaction instead of treating $\Dl$(1232) explicitly~\cite{horii14}.  The two-body interaction AV14 is written as the sum of many operators:
\beq
V=\fra \sum_{p=1}^{14} \sum_{i\ne j}V^p (r_{ij})~.
\label{interaction}
\eeq
There are 14 operators in the AV14 potential.  We write the two-body interaction as $\fra \sum_{i\ne j}$ instead of $\sum_{i<j}$ in order to treat the antisymmetrization systematically.
The Argonne potential has three radial components: a long-range one-pion exchange part $v_{\pi}$, and phenomenological intermediate-range and short-range core parts $v_{I}$ and $v_{S}$:
\beq
V^p(r_{ij})=\sum_{p}[ v_{\pi}^p (r_{ij}) + v_I^p(r_{ij}) + v_{S}^p(r_{ij})] \mathcal O_{ij}^p ~,
\label{int2}
\eeq
where operators $\mathcal O^p$ represent operators of spin, isospin, tensor, spin-orbit, squared angular momentum, and squared spin-orbit interactions.  In our calculations, we expand all the radial dependence in terms of the sum of Gaussian functions:
\beq
v_{\pi}^p (r_{ij}) + v_I^p(r_{ij}) + v_{S}^p(r_{ij})=\sum_\mu C^p_\mu\, e^{-a_\mu^p r_{ij}^2}~.
\eeq
As for the tensor operator, we multiply $r_{ij}^2$ in the radial dependence.

We have the three-body interaction in order to describe quantitatively the nuclear system:
\beq
U=\frac{1}{2} \sum_{i\ne j\ne k} (U^{2\pi}(ijk)+U^{R}(ijk))~.
\label{int3}
\eeq
There are two components for the three-body interaction in the Urbana series of three-nucleon potentials~\cite{pieper01}, where one term originates from two-pion exchange through delta excitation and the other term originates from the relativistic effect.  We write $\frac{1}{2} \sum_{i\ne j\ne k}$ instead of $\sum_{\rm cyclic}\sum_{i< j< k}U(ijk)=\sum_{i< j< k}(U(ijk)+U(jki)+U(kij))$ for systematic manipulations.  The two-pion exchange term is written as:
\beq
U^{2\pi}(ijk)=A_{2\pi}\{X_{ij}(r),X_{jk}(r)\}\{\tau_i\cdot \tau_j,\tau_j \cdot \tau_k\}+C_{2\pi}[X_{ij}(r),X_{jk}(r)][\tau_i\cdot \tau_j,\tau_j \cdot \tau_k]~.
\eeq
Here, all the operators with the radial dependences are those of the one-pion exchange:
\beq
X_{ij}(r)&=&Y(r)\sigma_i\cdot \sigma_j+T(r)S_{12}~,\nn
Y(r)&=&\frac{e^{-m_\pi r}}{m_\pi r}\xi_Y (m_\pi r)~,\nn
T(r)&=&\left(\frac{3}{(m_\pi r)^2}+\frac{3}{m_\pi r}+1\right)Y(r)\xi_T(r)~.
\eeq
Here, $\xi_Y$ and $\xi_T$ represent the short-range cut-off factors of these interactions~\cite{pieper01}.
We expand $Y(r)$ and $T(r)$ in the sum of Gaussian functions.
As for the relativistic effect term, the Argonne group chose the following expression:
\beq
U^R(ijk)=A_{R} T^2(m_\pi r_{ij})T^2(m_\pi r_{jk})~.
\eeq
This three-body interaction does not depend on spin and isospin.

\subsection{One-, two-, and three-body matrix elements}

We describe here how to calculate one-, two-, and three-body matrix elements.  We want to show all the necessary ingredients for matrix elements of the AMD wave function.  We give the matrix elements of the kinetic energy and those for the central, tensor, spin-orbit, squared angular momentum, and squared spin-orbit interactions, and for the three-body interactions.  

\vspace*{0.10cm}
\noindent\underline{\it Kinetic energy}\\
The matrix element of the individual kinetic energy for the AMD wave function is written as:
\beq
\bra \AMD|\sum_i\left(-\frac{\nabla_i^2}{2m}\right)|\AMD \ket
&=& \bra p_1 p_2 p_3\cdots |\sum_i\left(-\frac{\nabla_i^2}{2m}\right)| \det |q_1 q_2 q_3\cdots |\ket
\nonumber\\
&=&\hspace*{-0.2cm}
\sum_{pq}\bra p |-\frac{\nabla^2}{2m}| q \ket (B^{-1})_{qp}|B|=
\sum_{pq} \bra p |-\frac{\nabla^2}{2m}| q \ket C(p:q)~.
\eeq
Here, $C(p:q)=(B^{-1})_{qp} |B|$ is the determinant of a co-factor matrix for the $pq$ element of the overlap matrix $B$ and it is obtained using the properties of the determinant. All the details of the co-factor matrix $C$ are given in Appendix \ref{sec:matrix}.  Here, $(B^{-1})_{qp}$ denotes the $qp$ component of the inverse matrix of the overlap matrix $B$.  The single-particle matrix element is calculated in  rectangular coordinates for the AMD single-particle states:
\beq
\bra p |-\frac{\nabla^2}{2m}| q \ket=\frac{1}{2m}\left(3\nu-\nu^2(\vec D_p-\vec D_q)^2\right) e^{-\fra \nu(\vec D_p-\vec D_q)^2}M^{pq}\bar M^{pq}~.
\eeq
Here, $M^{pq}$ is the spin matrix element (\ref{spinm}) and $\bar M^{pq}$ the isospin matrix element (\ref{isospinm}).

As for the center-of-mass kinetic energy, the center-of-mass wave function of the AMD wave function is:
\beq
\Psi_{\rm c.m.}(\vec R)=\left(\frac{2\tilde \nu}{\pi}\right)^{3/4} e^{-\tilde \nu(\vec R-\tilde D)^2}~,
\eeq
where $\vec R$ is the center-of-mass coordinate, $\tilde \nu=A\nu$, and $\tilde D=\frac{1}{A}\sum_i^A \vec D_i$.  The center-of-mass kinetic energy is $\frac{-1}{2(Am)}\nabla_R^2$ and $\tilde D=0$ is assumed.  Hence, the matrix element is:
\beq
\bra \Psi_{\rm c.m.}|\frac{-1}{2(Am)}\nabla_R^2|\Psi_{\rm c.m.}\ket=\frac{3\tilde \nu}{2(Am)}=\frac{3\nu}{2m}
\label{kinsp}
\eeq
This is a single-particle kinetic energy.  Since the interactions and correlations are written in terms of the relative (intrinsic) coordinates and momenta, the center-of-mass state is unaffected.  Therefore, the intrinsic energy should be obtained after calculating everything using the individual kinetic energy and subtracting the center-of-mass kinetic energy (\ref{kinsp}).

\vspace*{0.10cm}
\noindent\underline{\it Central interaction}\\
The central interaction is expanded in the Gaussian functions:
\beq
V^c=\fra \sum_{i\ne j} \sum_\mu C_\mu e^{-a_\mu r_{ij}^2}~.
\eeq
In order to perform the antisymmetrization in a systematic manner (Appendix \ref{sec:matrix}), we write the interaction in the momentum representation so that the operator has a separable form.
\beq
V^c=\fra \sum_{i\ne j} \sum_\mu C_\mu \left(\frac{\pi}{a_\mu}\right)^{3/2}\int_k e^{- k^2/4a_\mu}e^{i\vec k \vec r_i}e^{-i\vec k\vec r_j}~.
\eeq
The momentum integration is written in a shorthand notation:
\beq
\label{mom}
\int _k=\int \frac{d^3 k}{(2\pi)^3}~.
\eeq
We can write the matrix element as:
\beq
&&\bra \AMD|V^c|\AMD \ket
=\fra \sum_{pqst} \sum_\mu C_\mu \left(\frac{\pi}{a_\mu}\right)^{3/2}
\nonumber\\
&& ~~\int_k e^{- k^2/4a_\mu} \bra p| e^{i\vec k \vec r_i}|s \ket \bra q| e^{-i\vec k \vec r_j} |t \ket \left((B^{-1})_{sp}(B^{-1})_{tq}-(B^{-1})_{tp}(B^{-1})_{sq}\right)|B|~.
\eeq
By writing the interaction in the momentum representation, we are able to handle the antisymmetrization systematically.  We give the antisymmetrization in terms of a co-factor matrix for the $pq:st$ element of the $B$ matrix: $C(pq:st)=\left((B^{-1})_{sp}(B^{-1})_{tq}-(B^{-1})_{tp}(B^{-1})_{sq}\right)|B|$.  We can calculate the single-particle matrix element easily using the Gaussian integration formula and write the final expression as:
\beq
\bra \AMD|V^c|\AMD \ket
=\fra \sum_\mu C_\mu \left(\frac{\pi}{a_\mu}\right)^{3/2} \sum_{pqst} I^{(12)}(A,B,C)\
M^{ps}M^{qt}\bar M^{ps}\bar M^{qt}\, C(pq:st)~.
\eeq
Here, the spatial matrix element is:
\beq
I^{(12)}(A,B,C)
&=& \int_k e^{-k^2/4a_\mu} \bra \psi_p|e^{i\vec{k}\vec{r}_i}|\psi_s\ket \bra \psi_q|e^{-i\vec{k}\vec{r}_j}|\psi_t\ket 
\nonumber\\
&=&
\frac{1}{(2\pi)^3}\left(\frac{\pi}{A}\right)^{3/2}e^{-\frac{ B^2}{4A}+C}~.
\label{fun1}
\eeq
All the coefficients $A, B, C$ in the Gaussian integral (\ref{fun1}) are given in Appendix \ref{sec:gauss}.
Since we have to handle many Gaussian integrals, we have introduced a notation for the Gaussian integrals of $I^{(type)}(A,B,C)$.
We introduce the superscript as $(type)$ for the integral, where $type$ denotes which nucleons interact with each other.  In the present example (\ref{fun1}), particles 1 and 2 interact with each other by a two-body interaction ($type=12$).  All the Gaussian integrals $I^{(type)}(A,B,C)$ up to three-body operators are given explicitly in Appendix \ref{sec:gauss}.  

In the case of spin-spin interaction $\sigma_1\cdot \sigma_2$, we have to replace the spin matrix elements $M^{ps}M^{qt}$ with $\sum_x M_x^{ps} M_x^{qt}$, where
\beq
M_x^{ps}=\bra \chi_p|\sigma_x|\chi_s\ket~.
\eeq
We calculate these spin matrices using the Pauli spin operators and the spin-spinor wave functions. 
In the case of isospin-isospin interaction $\tau_1\cdot \tau_2$, we have to replace the isospin matrix elements $\bar M^{ps}\bar M^{qt}$ with $\sum_x \bar M_x^{ps} \bar M_x^{qt}$, where
\beq
\bar M_x^{ps}=\bra \xi_p|\tau_x|\xi_s \ket~.
\eeq

\vspace*{0.10cm}
\noindent\underline{\it Tensor interaction}\\
The matrix element of the tensor interaction, where we see all the essential features of the tensor operator, is interesting.  The tensor interaction in the momentum space is written as:
\beq
&&V^t=\fra \sum_{i\ne j} \sum_\mu C_\mu \left(\frac{\pi}{a_\mu}\right)^{3/2}\left(\frac{-1}{4a^2_\mu}\right)\nonumber\\&&~~\sum_{xyx'y'}\int_{k_1}  e^{- k_1^2/4a_\mu}e^{i\vec k_1 \vec r_i}e^{-i\vec k_1 \vec r_j} k_{1x}k_{1y}\sigma_{ix'}\sigma_{jy'}(3\delta_{xx'}\delta_{yy'}-\delta_{xy}\delta_{x'y'})~.
\label{vten}
\eeq
The matrix element of the tensor interaction is written as:
\beq
&&\bra \AMD|V^t|\AMD \ket
=\fra \sum_{pqst} \sum_\mu C_\mu \left(\frac{\pi}{a_\mu}\right)^{3/2}\left(\frac{-1}{4a^2_\mu}\right)\nonumber\\&&~~ \sum_{xyx'y'} \int_{k_1} e^{- k_1^2/4a_\mu}k_{1x}k_{1y} (3\delta_{xx'}\delta_{yy'}-\delta_{xy}\delta_{x'y'})
\bra p| e^{i\vec k_1 \vec r_i}\sigma_{x'}|s \ket \bra q| e^{-i\vec k_1 \vec r_j}\sigma_{y'} |t \ket \nonumber\\&&~~\left((B^{-1})_{sp}(B^{-1})_{tq}-(B^{-1})_{tp}(B^{-1})_{sq}\right)|B|~. 
\label{tenmat}
\eeq
Using various matrix elements we can write it as:
\beq
&&\bra \AMD|V^t|\AMD \ket
=\fra \sum_{pqst} \sum_\mu C_\mu \left(\frac{\pi}{a_\mu}\right)^{3/2}\left(\frac{-1}{4a^2_\mu}\right) \sum_{xyx'y'} I_{1x1y}^{(12)}(A,B,C)\nonumber\\&&~~(3\delta_{xx'}\delta_{yy'}-\delta_{xy}\delta_{x'y'})M_{x'}^{ps}M_{y'}^{qt}\bar M^{ps}\bar M^{qt} C(pq:st)~.
\label{int1}
\eeq
Here, the momentum integration is defined as:
\beq
I_{1x1y}^{(12)}(A,B,C)=\int_{k_1}k_{1x}k_{1y}e^{-k^2/4a_\mu}\bra \psi_p|e^{i\vec{k}\vec{r}_i}|\psi_s\ket \bra \psi_q|e^{-i\vec{k}{\vec r}_j}|\psi_t\ket~,
\eeq
where the coefficients $A, B, C$ are given in Appendix \ref{sec:gauss}.  The subscript $1x1y$ is related to the momentum integrations for momenta $k_{1x}k_{1y}$ outside of the exponent (\ref{tenmat}) for multiple momentum integrations to be discussed in the next section.  If we have isospin-isospin interaction, we should replace $\bar M^{ps}\bar M^{qt}$ with $\sum_ x\bar M_x^{ps}\bar M_x^{qt}$.  Hereafter, we do not write this statement for other interactions.  Later we will introduce a matrix form for treating multiple tensor operator cases and describe a systematic method to calculate momentum integrations.  

\vspace*{0.10cm}
\noindent\underline{\it Spin-orbit interaction}\\
The spin-orbit interaction is written as:
\beq
V^{ls}=\fra \sum_{i\ne j} \sum_\mu C_\mu e^{-a_\mu r_{ij}^2} \vec l_{ij}\cdot (\sigma_i+\sigma_j)/2~,
\eeq
where the relative orbital angular momentum $\vec l_{ij}$ is written as $\vec l_{ij}=\vec r_{ij} \times \fra(\vec p_i-\vec p_j)$.
In the momentum space it is written as:
\beq
&&V^{ls}=\fra \sum_{i\ne j} \sum_\mu C_\mu \left(\frac{\pi}{a_\mu}\right)^{3/2}\left(\frac{-i}{2a_\mu}\right)\int_{k_1}e^{- k_1^2/4a_\mu}e^{i\vec k_1 \vec r_i}e^{-i\vec k_1\vec r_j}\nonumber\\&&~~\sum_{xyz} \varepsilon_{xyz} k_{1x} \fra(p_i-p_j)_y\fra(\sigma_{i}+\sigma_{j})_z ~.
\label{vspino}
\eeq
We need the matrix element of $p_x=-i\nabla_x$:
\beq
&&\bra \psi_p|e^{i\vec k_1\vec r}p_x e^{i\vec k_2\vec r} |\psi_q\ket\nonumber\\&&=\left(\fra (k_{2x}-k_{1x})+i\nu(D_p-D_q)_x\right) e^{-\fra \nu(D_p-D_q)^2+\fra i(k_1+k_2)(D_p+D_q)-(k_1+k_2)^2/8\nu}~.
\label{single}
\eeq
The matrix element of the spin-orbit interaction is written as:
\beq
&&\bra \AMD|V^{ls}|\AMD \ket
=\fra \sum_{pqst} \sum_\mu C_\mu \left(\frac{\pi}{a_\mu}\right)^{3/2}\left(\frac{-i}{2a_\mu}\right)\nonumber\\&&~~ \sum_{xyz} \varepsilon_{xyz} \int_{k_1} e^{- k_1^2/4a_\mu}k_{1x}\fra i\nu((D_p-D_s)-(D_q-D_t))_y\nonumber\\&&~~ \fra \left[\bra p|e^{ik_1r}\sigma_z|s\ket  \bra q|e^{-ik_1r}|t\ket+\bra p|e^{ik_1r}|s\ket  \bra q|e^{-ik_1r}\sigma_z|t\ket\right]  C(pq:st)~.
\eeq
Using various matrix elements we can write it as:
\beq
&&\bra \AMD|V^{ls}|\AMD \ket
=\fra \sum_{pqst} \sum_\mu C_\mu \left(\frac{\pi}{a_\mu}\right)^{3/2}\left(\frac{-i}{2a_\mu}\right) \sum_{xyz}\varepsilon_{xyz} I_{1x}^{(12)}(A,B,C)\nonumber\\&&~~  \fra i\nu((D_p-D_s)-(D_q-D_t))_y\fra(M_z^{ps}M^{qt}+M^{ps}M_z^{qt})\bar M^{ps}\bar M^{qt} C(pq:st)~.
\label{int2}
\eeq

\vspace*{0.10cm}
\noindent\underline{\it Squared angular momentum interaction}\\
The squared angular momentum interaction is written as:
\beq
&&V^{l^2}=\fra \sum_{i\ne j} \sum_\mu C_\mu e^{-a_\mu r_{ij}^{2}} \vec l_{ij}^{\,2}=\fra \sum_{i\ne j} \sum_\mu C_\mu e^{-a_\mu r_{ij}^{2}} (\vec r_{ij}\times \vec p_{ij})^2\nonumber\\&&=\fra \sum_{i\ne j} \sum_\mu C_\mu e^{-a_\mu r_{ij}^{2}} \left(\sum_{xyzuv}\varepsilon_{xyz}\varepsilon_{uvz} xup_{ijy}p_{ijv}+2i\sum_x xp_{ijx}\right)\nonumber\\&&=\fra \sum_{i\ne j} \sum_\mu C_\mu e^{-a_\mu r_{ij}^{2}} \frac{1}{4}\left(\sum_{xyzuvz}\hspace*{-0.2cm}\varepsilon_{xyz}\varepsilon_{uvz} xu(p_i-p_j)_y(p_i-p_j)_v+4i\sum_x x(p_i-p_j)_x\right)~.
\eeq
In the momentum space it is written as:
\beq
&&V^{l^2}=\fra \sum_{i\ne j} \sum_\mu C_\mu \left(\frac{\pi}{a_\mu}\right)^{3/2}\int_{k_1} e^{- k_1^2/4a_\mu}e^{i\vec k_1 \vec r_i}e^{-i\vec k_1\vec r_j}\nonumber\\&&~~ \frac{1}{4}\left[\left(\frac{-i}{2a_\mu}\right)^2\sum_{xyzuv} \varepsilon_{xyz}\varepsilon_{uvz}  (k_{1x}k_{1u}-2a_\mu\delta_{xu}) (p_i-p_j)_y(p_i-p_j)_v \right.\nonumber\\&&
\left. +\left(\frac{-i}{2a_\mu}\right)4i\sum_{x}k_{1x}(p_{i}-p_{j})_x\right]~.
\label{vang2}
\eeq
We write the matrix element of the double-differential operator $p_xp_y=-\nabla_x\nabla_y$:
\beq
&&\bra \psi_p|e^{i\vec k_1\vec r}p_x p_y e^{i\vec k_2\vec r} |\psi_q\ket=\Biggl[\nu\delta_{xy}+\left(\fra(k_{2x}-k_{1x})+i\nu(D_p-D_q)_x\right)\nonumber\\&&~~\left(\fra(k_{2y}-k_{1y})+i\nu(D_p-D_q)_y\right)\Biggr] 
e^{-\fra \nu(D_p-D_q)^2+\fra i(k_1+k_2)(D_p+D_q)-(k_1+k_2)^2/8\nu}~.
\label{double}
\eeq
The matrix element of the squared angular momentum interaction using the single (\ref{single}) and double derivatives (\ref{double}) is written as:
\beq
&&\bra \AMD|V^{l^2}|\AMD \ket
=\fra \sum_{pqst} \sum_\mu C_\mu \left(\frac{\pi}{a_\mu}\right)^{3/2} \left(\frac{-i}{2a_\mu}\right)^2 \frac{1}{4}\nonumber\\ && \left[\sum_{xyzuv}\varepsilon_{xyz} \varepsilon_{uvz}\left[I_{1x1u}^{(12)}(A,B,C)-2a_\mu\delta_{xu}I^{(12)}(A,B,C)\right] \right.\nonumber\\ &&
\Bigl[2\nu\delta_{yv} -\nu^2((D_p-D_s)-(D_q-D_t))_y((D_p-D_s)-(D_q-D_t))_v\Bigr] \nonumber\\&&\left. +4a_\mu \sum_x I_{1x1x}^{(12)}(A,B,C)\right]
M^{ps}M^{qt}\bar M^{ps} \bar M^{qt}\, C(pq:st)~.
\label{ls2}
\eeq
\vspace*{0.10cm}
\noindent\underline{\it Squared spin-orbit interaction}\\
The squared spin-orbit interaction is written as:
\beq
V^{(ls)^2}=\fra \sum_{i\ne j} \sum_\mu C_\mu e^{-a_\mu r_{ij}^2} (\vec l_{ij}\cdot (\sigma_i+\sigma_j)/2)^2~.
\eeq
In the momentum space it is written as:
\beq
&&V^{(ls)^2}=\fra \sum_{i\ne j} \sum_\mu C_\mu^{(2)} \left(\frac{\pi}{a_\mu}\right)^{3/2} \left(\frac{-i}{2a_\mu}\right)^2\int_{k_1} e^{- k_1^2/4a_\mu}e^{i\vec k_1 \vec r_i}e^{-i\vec k_1\vec r_j}\nonumber\\&&~~ \frac{1}{16}\left[\sum_{xyzuvw} \varepsilon_{xyz}\varepsilon_{uvw}  (k_{1x}k_{1u}-2a_\mu\delta_{xu}) (p_i-p_j)_y(p_i-p_j)_v \right.\nonumber\\&&
\left. +4a_\mu \sum_{xyzuvw} \varepsilon_{xyz}\varepsilon_{uvw} \delta_{yu}k_{1x}(p_{i}-p_{j})_v\right](\sigma_{i}+\sigma_{j})_z(\sigma_{i}+\sigma_{j})_w~.
\label{vspino2}
\eeq
The matrix element of the squared spin-orbit interaction is written as:
\beq
&&\bra \AMD|V^{(ls)^2}|\AMD \ket
=\fra \sum_{pqst} \sum_\mu C_\mu \left(\frac{\pi}{a_\mu}\right)^{3/2}\left(\frac{-i}{2a_\mu}\right)^2 \frac{1}{16}\nonumber\\ &&\sum_{xyzuvw}\varepsilon_{xyz} \varepsilon_{uvw}\biggl[ [I_{1x1u}^{(12)}(A,B,C)-2a_\mu\delta_{xu}I^{(12)}(A,B,C)]\nonumber\\ &&\Bigl[2\nu\delta_{yv} -\nu^2((D_p-D_s)-(D_q-D_t))_y((D_p-D_s)-(D_q-D_t))_v\Bigr]  -2a_\mu  \delta_{yu}I_{1x1v}^{(12)}(A,B,C)\biggr]\nonumber\\ &&
(M_{zw}^{ps}M^{qt}+M_z^{ps}M_w^{qt}+M_w^{ps}M_z^{qt}+M^{ps}M_{zw}^{qt})\bar M^{ps} \bar M^{qt}\, C(pq:st)~.
\label{ls2}
\eeq

\vspace*{0.10cm}
\noindent\underline{\it Short-range three-body interaction}\\
The three-body interaction $U$ contains two components, one due to two-pion exchange exciting the $\Delta$ state and one due to the relativistic effect. For calculation of the matrix element it is simpler to give the relativistic-effect one first.
The short-range three-body interaction due to the relativistic effect is written as:
\beq
U^{R}= \sum_{i\ne j\ne k} \left(\fra A_R\right) T^2(r_{ij})T^2(r_{jk})= \sum_{i\ne j \ne k} \left(\fra A_R\right)\sum_{\mu1} C_{\mu1} e^{-a_{\mu1} r_{ij}^2} \sum_{\mu2} C_{\mu2} e^{-a_{\mu2} r_{jk}^2}~.
\eeq
Hence, the matrix element is
\beq
&&\bra \AMD|U^{R}|\AMD \ket=\sum_{pqrstu}\left(\fra A_R\right) \sum_{\mu1} C_{\mu1}\left(\frac{\pi}{a_{\mu1}}\right)^{3/2}\sum_{\mu2} C_{\mu2} \left(\frac{\pi}{a_{\mu2}}\right)^{3/2}\nonumber\\ &&~~\int_{k_1}\int_{k_2}  e^{- k_1^2/4a_{\mu1}}e^{- k_2^2/4a_{\mu2}}\bra p|e^{i\vec k_1 \vec r}|s\ket\bra q|e^{i(-\vec k_1+\vec k_2)\vec r}|t\ket \bra r|e^{-i\vec k_2 \vec r}|u\ket C(pqr:stu)\nonumber\\ &&~~=\sum_{pqrstu}\left(\fra A_R\right) \sum_{\mu1} C_{\mu1} \left(\frac{\pi}{a_{\mu1}}\right)^{3/2}\sum_{\mu2} C_{\mu2}\left(\frac{\pi}{a_{\mu2}}\right)^{3/2}I^{(12:23)}(A,B,C) \nonumber\\ && ~~M^{ps}M^{qt}M^{ru}\bar M^{ps}\bar M^{qt}\bar M^{ru}\, C(pqr:stu)~.
\label{three}
\eeq
Here, $C(pqr:stu)$ is the determinant of the co-factor matrix for the $pqr:stu$ matrix of the overlap matrix $B$ as explained in Appendix A.  The explicit form of the Gaussian integral $I^{(12:23)}$ is written in Appendix B, where $(type=12:23)$ indicates that the first two-body operator acts on particles 1 and 2 and the second two-body operator on particles 2 and 3.

\vspace*{0.10cm}
\noindent\underline{\it Two-pion three-body interaction}\\
There are two terms for the two-pion three-body interaction, 
$\{X_{ij},X_{jk}\}\{\tau_i\tau_j,\tau_j\tau_k\}$ and $[X_{ij},X_{jk}][\tau_i\tau_j,\tau_j\tau_k]$.  First we decompose $\{X_{ij},X_{jk}\}$ in the spin and tensor operators:
\beq
&&X_{ij}X_{jk}+X_{jk}X_{ij}\nonumber\\ &&~~=
Y(r_{ij})Y(r_{jk})(\sigma_i\cdot \sigma_j\sigma_j\cdot\sigma_k+\sigma_j\cdot \sigma_k\sigma_i\cdot\sigma_j)\nonumber\\ &&~~+
T(r_{ij})T(r_{jk})(S_{12}(ij)S_{12}(jk)+S_{12}(jk)S_{12}(ij))\nonumber\\ &&~~+
Y(r_{ij})T(r_{jk})(\sigma_i\cdot\sigma_jS_{12}(jk)+S_{12}(jk)\sigma_i\cdot\sigma_j)\nonumber\\ &&~~+
Y(r_{jk})T(r_{ij})(S_{12}(ij)\sigma_j\cdot\sigma_k+\sigma_j\cdot\sigma_k S_{12}(ij))~.
\label{st3}
\eeq
These spatial-spin operators are multiplied by the isospin operators.
\beq
\tau_i \cdot \tau_j \tau_j \cdot \tau_k+\tau_j \cdot \tau_k \tau_i \cdot \tau_j~.
\eeq

We start with the spin-spin three-body part:
\beq
&&U^{2\pi:spin-spin}=\left(\fra A_\pi\right) \sum_{i\ne j\ne k} Y(r_{ij})Y(r_{jk})(\sigma_i\cdot \sigma_j\sigma_j\cdot\sigma_k+\sigma_j\cdot \sigma_k\sigma_i\cdot\sigma_j)\nonumber\\ &&~~(\tau_i \cdot \tau_j \tau_j \cdot \tau_k+\tau_j \cdot \tau_k \tau_i \cdot \tau_j)\nonumber\\ &&~~= \sum_{i\ne j \ne k}\left(\fra A_\pi\right) \sum_{\mu1} C_{\mu1}  e^{-a_{\mu1} r_{ij}^2} \sum_{\mu2} C_{\mu2} e^{-a_{\mu2} r_{jk}^2}(\sigma_i\cdot \sigma_j\sigma_j\cdot\sigma_k+\sigma_j\cdot \sigma_k\sigma_i\cdot\sigma_j)\nonumber\\ &&~~(\tau_i \cdot \tau_j \tau_j \cdot \tau_k+\tau_j \cdot \tau_k \tau_i \cdot \tau_j)~.
\eeq
Hence, the matrix element is:
\beq
&&\bra \AMD|U^{2\pi:spin-spin}|\AMD \ket=\sum_{pqrstu} \left(\fra A_\pi\right) \sum_{\mu1} C_{\mu1}\left(\frac{\pi}{a_{\mu1}}\right)^{3/2}\sum_{\mu2} C_{\mu2} \left(\frac{\pi}{a_{\mu2}}\right)^{3/2}\nonumber\\ &&~~ \sum_{xy} I^{(12:23)}(A,B,C) (M_x^{ps}M_{xy}^{qt}M_y^{ru}+M_y^{ps}M_{xy}^{qt}M_x^{ru})\nonumber\\ &&~~\sum_{vw}(\bar M_{v}^{ps}\bar M_{vw}^{qt}\bar M_{w}^{ru}+\bar M_{w}^{ps}\bar M_{vw}^{qt}\bar M_{v}^{ru})\ C(pqr:stu)~.
\label{tp1}
\eeq

We write the tensor-tensor three-body part:
\beq
&&U^{2\pi:tensor-tensor}=\left(\fra A_\pi\right) \sum_{i\ne j\ne k} T(r_{ij})T(r_{jk})(S_{12}(ij)S_{12}(jk)+S_{12}(jk)S_{12}(ij))\nonumber\\ &&~~~~~~(\tau_i \cdot \tau_j \tau_j \cdot \tau_k+\tau_j \cdot \tau_k \tau_i \cdot \tau_j)\nonumber\\ &&~~= \sum_{i\ne j \ne k}\left(\fra A_\pi\right) \sum_{\mu1} C_{\mu1} r_{ij}^2e^{-a_{\mu1} r_{ij}^2} \sum_{\mu2} C_{\mu2} r_{jk}^2 e^{-a_{\mu2} r_{jk}^2}(S_{12}(ij)S_{12}(jk)+S_{12}(jk)S_{12}(ij))\nonumber\\ &&~~~~~~(\tau_i \cdot \tau_j \tau_j \cdot \tau_k+\tau_j \cdot \tau_k \tau_i \cdot \tau_j)~.
\eeq
Hence, the matrix element is:
\beq
&&\bra \AMD|U^{2\pi:tensor-tensor}|\AMD \ket= \sum_{pqrstu} \left(\fra A_\pi\right) \sum_{\mu1} C_{\mu1} \left(\frac{\pi}{a_{\mu1}}\right)^{3/2}\left(\frac{-1}{4a_{\mu1}^2}\right) \nonumber\\ &&~~\sum_{\mu2} C_{\mu2}\left(\frac{\pi}{a_{\mu2}}\right)^{3/2}\left(\frac{-1}{4a_{\mu2}^2}\right) \sum_{xyzu\,x'y'z'u'} I_{1x1y2z2u}^{(12:23)}(A,B,C) \nonumber\\ &&~~(3\delta_{xx'}\delta_{yy'}-\delta_{xy}\delta_{x'y'})(3\delta_{zz'}\delta_{uu'}-\delta_{zu}\delta_{z'u'})(M_{x'}^{ps}M_{y'z'}^{qt}M_{u'}^{ru}+M_{z'}^{ps}M_{x'u'}^{qt}M_{y'}^{ru})\nonumber\\ &&~~\sum_{vw}(\bar M_v^{ps}\bar M_{vw}^{qt}\bar M_w^{ru}+\bar M_w^{ps}\bar M_{vw}^{qt}\bar M_v^{ru})C(pqr:stu)~.
\label{tp2}
\eeq

We write the spin-tensor three-body part:
\beq
U^{2\pi:spin-tensor}&=&2\left(\fra A_\pi\right) \sum_{i\ne j\ne k} Y(r_{ij})T(r_{jk})(\sigma_i\cdot \sigma_j S_{12}(jk)+S_{12}(jk)\sigma_i\cdot \sigma_j)\nonumber\\ 
&&(\tau_i \cdot \tau_j \tau_j \cdot \tau_k+\tau_j \cdot \tau_k \tau_i \cdot \tau_j)~.
\eeq
Here, because of the change in the particle coordinates, the third and fourth terms in Eq.\,(\ref{st3}) are identical and we multiply by 2 for the spin-tensor three-body term.
Hence, the matrix element is:
\beq
&&\bra \AMD|U^{2\pi:spin-tensor}|\AMD \ket= 2\sum_{pqrstu} \left(\fra A_\pi\right) \sum_{\mu1} C_{\mu1} \left(\frac{\pi}{a_{\mu1}}\right)^{3/2} \nonumber\\ &&~~\sum_{\mu2} C_{\mu2}\left(\frac{\pi}{a_{\mu2}}\right)^{3/2}\left(\frac{-1}{4a_{\mu2}^2}\right) \sum_{xyx'y'}\sum_{z'} I_{2x2y}^{(12:23)}(A,B,C) \nonumber\\ &&~~(3\delta_{xx'}\delta_{yy'}-\delta_{xy}\delta_{x'y'})(M_{z'}^{ps}M_{z'x'}^{qt}M_{y'}^{ru}+M_{z'}^{ps}M_{x'z'}^{qt}M_{y'}^{ru})\nonumber\\ &&~~\sum_{vw}(\bar M_v^{ps}\bar M_{vw}^{qt}\bar M_w^{ru}+\bar M_w^{ps}\bar M_{vw}^{qt}\bar M_v^{ru})C(pqr:stu)~.
\label{tp3}
\eeq

As for the commutator terms:
\beq
[X_{ij},X_{jk}][\tau_i\tau_j,\tau_j\tau_k]=(X_{ij}X_{jk}-X_{jk}X_{ij})(\tau_i \cdot \tau_j \tau_j \cdot \tau_k-\tau_j \cdot \tau_k \tau_i \cdot \tau_j)~,
\eeq
the additions for the spin part of the above matrix elements ((\ref{tp1}), (\ref{tp2}), (\ref{tp3})) are replaced by subtractions and the additions for the isospin part are also replaced by subtractions.

\section{Two-body interaction with one tensor correlation operator}\label{sec:one_fd}

We give here the transition matrix element from the AMD wave function $|\AMD\ket$ to the AMD wave function with tensor correlation $F_D|\AMD\ket$.  This matrix element includes the overlap matrix of the tensor-correlated state of the TOAMD state.
\beq
\bra \AMD| V F_D| \AMD \ket~.
\eeq
Here, $V$ is given as a summation over the particle coordinates in Eq.\,(\ref{interaction}) and $F_D$ is also given as a summation over the particle coordinates in Eq.\,(\ref{cten}).  Hence, there are various many-body operators:
\beq
V F_D &=& \left(\fra \sum_{i\ne j} V_{ij}\right)\left( \fra \sum_{k\ne l} F_{kl}\right)= \fra \sum_{i\ne j}V_{ij}F_{ij}+\sum_{i\ne j\ne k}V_{ij}F_{jk}+\frac{1}{4}\sum_{i\ne j\ne k\ne l}V_{ij}F_{kl}\nonumber\\ &=&Q_2+Q_3+Q_4~.
\label{qn}
\eeq
Here, the symmetry factors $S$ in front of each term in $Q_n$ such as $\fra$, 1, and $\frac{1}{4}$ are obtained by taking into account the symmetry of interchange of the particle coordinates.
$S$ is tabulated in Table \ref{table} of Sect. 6.  Hence, there appear two-body, three-body and four-body operators, which are written as $Q_2$, $Q_3$, and $Q_4$.  We shall discuss the matrix elements of these operators one by one.

\vspace*{0.10cm}
\noindent\underline{\it Two-body term}\\
We write the case of the tensor interaction explicitly.  As for the central interaction, the expressions are similar:
\beq
Q_2=\fra \sum_{i\ne j} \left\{\sum_{\mu1} C_{\mu1} r_{ij}^2e^{-a_{\mu1}r_{ij}^2}  S_{12}(r_{ij})\right\} 
\left\{ \sum_{\mu2} C_{\mu2} r_{ij}^2e^{-a_{\mu2}r_{ij}^2}S_{12}(r_{ij})\tau_i\cdot\tau_j \right\}~.
\label{vten2}
\eeq
Here, the first bracket corresponds to the tensor interaction and the second bracket the tensor correlation. 
Each tensor operator is expressed in the momentum space using the following expression:
\beq
\sum_{\mu} C_{\mu} r_{ij}^2e^{-a_{\mu}r_{ij}^2}S_{12}(r_{ij})&=& \sum_\mu C_\mu \left(\frac{\pi}{a_\mu}\right)^{3/2}\left(\frac{-1}{4a^2_\mu}\right)\int_k e^{- k^2/4a_\mu}e^{i\vec k \vec r_i}e^{-i\vec k\vec r_j}\nonumber\\ &&\sum_{xyx'y'}  k_x k_y\sigma_{ix'}\sigma_{jy'}(3\delta_{xx'}\delta_{yy'}-\delta_{xy}\delta_{x'y'})~.
\eeq
From here we introduce a shorthand notation for the coefficient:
\beq
\tilde C_\mu^{(m)}=C_\mu \left(\frac{\pi}{a_\mu}\right)^{3/2}\left(\frac{-i}{2a_\mu}\right)^{m}~.
\eeq
The matrix element of the tensor interaction is written as:
\beq
&&\bra \AMD|Q_2|\AMD \ket
=\fra \sum_{pqst}\sum_{\mu1\mu2} \tilde C_{\mu1}^{(2)}\tilde C_{\mu2}^{(2)} \nonumber\\ &&~~\int_{k_1} \int_{k_2} e^{-k_1^2/4a_{\mu1}}e^{-k_2^2/4a_{\mu2}}
\bra \psi_p |e^{i(k_1+k_2)r_i} |\psi_s\ket \bra \psi_q|e^{-i(k_1+k_2) r_j} |\psi_t \ket 
\nonumber\\ &&~~\sum_{xyzu\,x'y'z'u'} k_{1x} k_{1y}k_{2z} k_{2u}
(3\delta_{xx'}\delta_{yy'}-\delta_{xy} \delta_{x'y'})(3\delta_{zz'}\delta_{uu'}-\delta_{zu}\delta_{z'u'}) \nonumber\\ &&~~M_{x'z'}^{ps}M_{y'u'}^{qt} \sum_w \bar M_w^{ps}\bar M_w^{qt} C(pq:st)~.
\eeq

Since we can write the single-particle matrix element,
\beq
\bra \psi_p|e^{i\vec k\vec r}|\psi_s\ket
=e^{-\fra\nu (\vec D_p-\vec D_s)^2+\fra i\vec k(\vec D_p+\vec D_s)-k^2/8\nu}~,
\eeq
in Gaussian form, we are left with integrations over momentum for the entire matrix element.  We describe the systematic method of the Gaussian integrals in the next section and in Appendix \ref{sec:gauss}, we give the final expression for the tensor matrix element:
\beq
&&\bra \AMD|Q_2|\AMD \ket
=\fra \sum_{pqst}\sum_{\mu1\mu2} \tilde C_{\mu1}^{(2)}\tilde C_{\mu2}^{(2)}
\sum_{xyzu\,x'y'z'u'}I_{1x1y2z2u}^{((12)^2)}(A,B,C)\nonumber\\ &&~~
(3\delta_{xx'}\delta_{yy'}-\delta_{xy}\delta_{x'y'})(3\delta_{zz'}\delta_{uu'}-\delta_{zu}\delta_{z'u'})M_{x'z'}^{ps}M_{y'u'}^{qt}\sum_w \bar M_w^{ps}\bar M_w^{qt}\ C(pq:st)~.
\label{int2}
\eeq
Here, the function $I_{1x1y2z2u}^{((12)^2)}(A,B,C)$ is the result of the momentum integration, which will appear in the next section.  The explicit form for $A, B, C$ is given in Appendix B.

\vspace*{0.10cm}
\noindent\underline{\it Three-body term}\\
The $Q_3$ operator is written as:
\beq
Q_3=\sum_{i\ne j\ne k} 
\left\{ \sum_{\mu1} C_{\mu1} r_{ij}^2e^{-a_{\mu1}r_{ij}^2}  S_{12}(r_{ij}) \right\}
\left\{ \sum_{\mu2} C_{\mu2} r_{jk}^2e^{-a_{\mu2}r_{jk}^2}S_{12}(r_{jk})\tau_j\cdot\tau_k \right\}~.
\label{vten3}
\eeq
The $Q_3$ operator corresponds to the case of $type=12:23$ for the momentum integration in the specification of Appendix B.  Hence, we can write the matrix element as:
\beq
&&\bra \AMD|Q_3|\AMD \ket
=\sum_{pqrsto}\sum_{\mu1\mu2} \tilde C_{\mu1}^{(2)}\tilde C_{\mu2}^{(2)} \sum_{xyzu\,x'y'z'u'} I_{1x1y2z2u}^{(12:23)}(A,B,C)\nonumber\\ &&~~(3\delta_{xx'}\delta_{yy'}-\delta_{xy} \delta_{x'y'})(3\delta_{zz'}\delta_{uu'}-\delta_{zu}\delta_{z'u'})M_x^{ps}M_{yz}^{qt}M_{u}^{ro}\nonumber\\ &&~~\bar M^{ps}\sum_w \bar M_w^{qt}\bar M_w^{ro}\ C(pqr:sto)~.
\eeq

\vspace*{0.10cm}
\noindent\underline{\it Four-body term}\\
The $Q_4$ operator is written as:
\beq
Q_4=\frac{1}{4} \sum_{i\ne j\ne k\ne l} 
\left\{ \sum_{\mu1} C_{\mu1} r_{ij}^2e^{-a_{\mu1}r_{ij}^2}  S_{12}(r_{ij}) \right\}
\left\{ \sum_{\mu2} C_{\mu2} r_{kl}^2e^{-a_{\mu2}r_{kl}^2}S_{12}(r_{kl})\tau_k\cdot\tau_l \right\} ~.
\label{vten4}
\eeq
The momentum integration of this four-body operator should be that of $type=12:34$ in Appendix B.  Hence, we are able to write the final result as:
\beq
&&\bra \AMD|Q_4|\AMD \ket
=\frac{1}{4}\sum_{pqrs\,p'q'r's'}\sum_{\mu1\mu2} \tilde C_{\mu1}^{(2)}\tilde C_{\mu2}^{(2)} \sum_{xyzu\,x'y'z'u'} I_{1x1y2z2u}^{(12:34)}(A,B,C)\nonumber\\ &&~~(3\delta_{xx'}\delta_{yy'}-\delta_{xy} \delta_{x'y'})(3\delta_{zz'}\delta_{uu'}-\delta_{zu}\delta_{z'u'})M_{x'}^{pp'}M_{y'}^{qq'}M_{z'}^{rr'}M_{u'}^{ss'}\nonumber\\ &&~~\bar M^{pp'}\bar M^{qq'}\sum_w \bar M_w^{rr'}\bar M_w^{ss'}C(pqrs:p'q'r's')~.
\eeq
We are able to write the momentum integral $I_{1x1y2z2u}^{(12:34)}(A,B,C)$ in a separable form and simplify the calculation of the matrix element.
If we take an interaction other than the tensor interaction, we should change the above formula slightly depending on the type of operators.

\section{Momentum integration and systematic differentiation}\label{sec:momentum}

As we have seen in the calculation of matrix elements, we have to perform various types of momentum integrations ($k_1,\ldots,k_l$) with Gaussian functions in the following form:
\beq
I_{ixjy\cdots kz}(A,B,C: b)=\int_{k_1} \int_{k_2}\cdots\int_{k_l} 
k_{ix}k_{jy}\cdots k_{kz}\ e^{-\vec k A \vec k+i\vec B \vec k+C}~.
\eeq
We have already used the cases where the number of the momentum integration is 1 in Eq.\,(\ref{int1}) and 2 in Eq.\,(\ref{int2}).
Here, a vector $\vec k$, a matrix $A$ and other quantities are defined as:
\beq
\vec k^t=\left( \begin{array}{cccc} \vec k_1 & \vec k_2 & \cdots & \vec k_l \\
\end{array} \right)
\eeq
\beq
A=\left( \begin{array}{cccc} 
1/4\nu+1/4a_{\mu 1} & A_{12} & \cdots & A_{1l}  \\
A_{21} & 1/4\nu+1/4a_{\mu 2} & \cdots & A_{2l}  \\
\vdots & \vdots & \ddots & \vdots  \\
A_{l1} & A_{l2} & \cdots & 1/4\nu+1/4a_{\mu l}  \\
\end{array} \right)
\eeq
Here, $A_{ij}$ is a fraction of $1/\nu$ and depends on the type of momentum integrals.
\beq
\vec B=\left( \begin{array}{c} 
\fra((\vec D_p+\vec D_{p'})-(\vec D_q+\vec D_{q'}))+\vec b_1  \\
\fra((\vec D_r+\vec D_{r'})-(\vec D_s+\vec D_{s'}))+\vec b_2  \\
\vdots  \\
\fra((\vec D_t+\vec D_{t'})-(\vec D_u+\vec D_{u'}))+\vec b_l  \\
\end{array} \right)
\eeq
and
\beq
C=-\fra \nu((D_p-D_{p'})^2+(D_q-D_{q'})^2+ \cdots +
(D_t-D_{t'})^2+(D_u-D_{u'})^2)~.
\eeq
By construction, the matrix $A$ is a real symmetric matrix and can be written as $A=A^{1/2}A^{1/2}$.
Again, the subscripts as $p, q\cdots$ of the ${D_p}'$s depend on the type of momentum integrals.
Here, we have included the source terms whose coefficients are written as $\vec b_1, \vec b_2,\ldots,\vec b_l$.  
The source terms provide the term $e^{i\vec b_i\vec k_i}$ in the integrand.  These source terms are used for the calculation of integrals of the form:
\beq
I_{ixjy\cdots kz}(A,B,C: b)
=\left(-i\frac{\partial}{\partial b_{ix}}\right) \left(-i\frac{\partial}{\partial b_{jy}}\right)\cdots\left(-i\frac{\partial}{\partial b_{kz}}\right)I(A,B,C:b)~.
\eeq
Here, $i, j,\ldots, k ~(\le l)$ represent the momenta and $b$ stands for the set of $\vec{b}_i$.  The momentum directions are denoted by $x, y,\ldots,z$.  We shall obtain the final results by setting the coefficients of the source terms $b$ to zero and write:
\beq
I_{ixjy\cdots kz}(A,B,C)=I_{ixjy\cdots kz}(A,B,C:b=0)~.
\eeq
These results of the momentum integrations are written in terms of the AMD wave functions with $\nu$ and various ${D_p}'$s and the interaction ranges $a_\mu$. 

We calculate $I_{ixiy\cdots kz}$ one by one.  In the following, $n$ denotes the number of momenta in the integration multiplied by the Gaussian functions.

\vspace*{0.10cm}
\noindent\underline {$n=0$}\\
We first take the integration of the basic integral:
\beq
I(A,B,C:b)&=&\int_{k_1} \int_{k_2}\cdots\int_{k_l} 
e^{-\vec k A \vec k+i\vec B \vec k+C}\nonumber\\ 
&=&\frac{1}{(2\pi)^{3 l}} \left(\frac{\pi^l}{\det|A|}\right)^{3/2}e^{-B^\dagger A^{-1}B/4+C}~.
\eeq
This multiple Gaussian integration is verified for a symmetric matrix $A$ with the existence of a square root matrix $A^{1/2}$.  The front factor $\frac{1}{(2\pi)^{3 l}}$ comes from the definition of the momentum integration in Eq.\,(\ref{mom}).  $\det|A|$ in the denominator appears from the Jacobian of the change of the integration variables and $\det |A^{-1/2}|=\frac{1}{\sqrt{\det|A|}}$ is used. 

\vspace*{0.10cm}
\noindent\underline {$n=1$}\\
\beq
I_{ix}(A,B,C:b)&=&\int_{k_1} \int_{k_2}\cdots\int_{k_l} k_{ix}
e^{-\vec k A \vec k+i\vec B \vec k+C}\nonumber\\ 
&=&-i\frac{\delta}{\delta b_{ix}}\int_{k_1} \int_{k_2}\cdots\int_{k_l} 
e^{-\vec k A \vec k+i\vec B \vec k+C}\nonumber\\ 
&=& -i\frac{\delta}{\delta b_{ix}}\frac{1}{(2\pi)^{3 l}} \left(\frac{\pi^l}{\det|A|}\right)^{3/2}e^{-B^\dagger A^{-1}B/4+C}~.
\eeq
Here, $b_{ix}$ is included in $\vec B$ and the derivative is:
\beq
-i\frac{\delta}{\delta b_{ix}}I(A,B,C:b)=iE_{ix}I(A,B,C:b)~,
\eeq
where
\beq
E_{ix}=(\del_{b_{ix}}B^\dagger)A^{-1}B_x/2=\fra \sum_j A^{-1}_{ij}B_{jx}~.
\eeq
Here, a very interesting relation is:
\beq
\del_{b_{ix}}B^\dagger=\left( \begin{array}{ccccccc}
0 & \cdots & 0 & 1 & 0 & \cdots & 0\\
\end{array}\right)~,
\eeq
where $1$ appears at the $i$ th position in the above vector of dimension $l$.

\vspace*{0.10cm}
\noindent\underline {$n=2$}\\
\beq
I_{ixjy}(A,B,C:b)&=&\int_{k_1} \int_{k_2}\cdots\int_{k_l} k_{ix}k_{jy}
e^{-\vec k A \vec k+i\vec B \vec k+C}\nonumber\\ 
&=&-i\frac{\delta}{\delta b_{jy}}\Bigl( iE_{ix}I(A,B,C:b) \Bigr) \nonumber\\ 
&=&(D_{ij}\delta_{xy}+iE_{ix}iE_{jy})I(A,B,C:b)~,
\eeq
where
\beq
D_{ij}\delta_{xy}=-i\frac{\partial}{\partial b_{jy}}iE_{ix}=\fra A^{-1}_{ij}\delta_{xy}~.
\eeq
Here, we have a symmetry in that the results are unchanged by changing the order of $ix$ and $jy$:
\beq
I_{jyix}(A,B,C:b)=I_{ixjy}(A,B,C:b)~.
\eeq

\vspace*{0.10cm}
\noindent\underline {$n=3$}\\
\beq
I_{ixjykz}(A,B,C:b)&=&\int_{k_1} \int_{k_2}\cdots\int_{k_l} k_{ix}k_{jy}k_{kz}
e^{-\vec k A \vec k+i\vec B \vec k+C}\nonumber\\ 
&=&-i\frac{\delta}{\delta b_{kz}}\Bigl\{ (D_{ij}\delta_{xy}+iE_{ix}iE_{jy})I(A,B,C:b) \Bigr\} \nonumber\\ 
&=&(D_{ij}\delta_{xy}iE_{kz}+D_{ik}\delta_{xz}iE_{jy}+D_{jk}\delta_{yz}iE_{ix}+iE_{ix}iE_{jy}iE_{kz})\nonumber\\ 
&\times& I(A,B,C:b)~.
\eeq
Here, we have used the fact that the derivative of $D_{ij}\delta_{xy}$ is zero:
\beq
-i\frac{\partial}{\partial b_{kz}}D_{ij}\delta_{xy}=0~.
\eeq

Since the derivative is done successively, we find several interesting rules found by deriving these expressions, which are useful for derivation of higher-order derivative terms:
\begin{itemize}
\item The derivative terms are written in terms of only $D_{ij}\delta_{xy}$ and $iE_{ix}$.  Because of this fact, we simply write $D_{ij}\delta_{xy}\rightarrow D_{\alpha\beta}$ and $iE_{ix}\rightarrow E_\alpha$ to express the derivative formula.
\item These $\alpha$ and $\beta$ denote $ix$ etc. At the same time, they can also mean the successive order of derivatives: $\alpha<\beta<\cdots$.
\end{itemize}
It is then interesting to write the properties of the differentiation using the numbering notation $I_{\alpha\beta\cdots}(A,B,C:b)$ using the above results up to $n=3$.  $I_{\alpha\beta\cdots}$ consists of the sum of $D_{\alpha \beta}^k(E_{\gamma}^{n-2k})$ terms with $k=1\cdots[\frac{n}{2}]$.  $I_{\alpha\beta\cdots}$ is symmetric for any exchange of the order of $\alpha, \beta,\ldots$ due to the interchangeable property of the differentiation.  Observing the terms with subscript $\alpha$ and the procedure of the above manipulations, the first subscript $\alpha$ appears once in all the terms keeping its position at the beginning for each term.

Hence, we can formally write the integrals as:
\beq
I_{\alpha\beta\gamma\cdots}(A,B,C:b)=\sum_{k=0}^{[\frac{n}{2}]}\sum_{q=1}^{N_k^n}[D^kE^{n-2k}]_{Q_q^k}I(A,B,C:b)~,
\eeq
where $[\frac{n}{2}]$ is $n/2$ for even $n$ and $(n-1)/2$ for odd $n$.
$Q_q^k$ denotes configurations of all the derivative terms $\alpha\beta\gamma\cdots$ to appear in the $D^k$ and $E^{n-2k}$ terms.  $N_k^n$ is the number of all the terms, where all $\alpha\beta\gamma\cdots$ are partitioned in the $D^k$ and $E^{n-2k}$ terms.

In order to write possible partitions of $\alpha\beta\gamma\cdots$, we consider the term $D^kE^{n-2k}$ and write rules for the configurations.  We write the configurations as:
\beq
[D^kE^{n-2k}]_{Q_q^k}=D_{a_1 b_1}D_{a_2 b_2} \cdots E_{c_1}E_{c_2} \cdots
\eeq
Here, $\alpha\beta\gamma\cdots$ are written in the form of $a_i b_j c_k\cdots$.  Given the observations above for the exchange property of $\alpha\beta\cdots$ and the order of the appearance of $\alpha$, we can write the following rules:
\begin{itemize}
\item Rule 1: $a_1 < b_1$, $a_2 < b_2$, $\ldots$
\item Rule 2: $a_1 <a_2 < \cdots $
\item Rule 3: $c_1 < c_2 < \cdots $
\end{itemize}
With these rules, we can write all the configurations without double counting of them. It is interesting to calculate the number of terms for each partition $N_k^n$.  There are altogether $n!$ ways to order the derivatives $\alpha\beta\gamma\cdots$.  There are three rules to avoid double counting of partitions in the $n!$ ways.  Now, the order of $c_i$ in $E^{n-2k}$ is fixed to a unique one from rule 3.  There are $(n-2k)!$ ways to order the derivatives, but the order is fixed to one by the rule 3.  Hence, we have to divide by $(n-2k)!$ out of the entire possibility $n!$.  We can have $k$ pairs, we have to divide by $2^k$ from rule 1.  In addition, there are $k$ $D's$; we have to order these $D's$ using rule 2 and we have to divide by $k!$.  Hence the number of configurations for each term is:
\beq
N_k^n=\frac{n!}{2^k k! (n-2k)!}~.
\eeq
For $n=3$ and $k=1$ it is $N_1^3=3$, and for $k=0$ it is $N_0^3=1$.  These numbers of terms agree with the above results.

Using the above rules, we write the $n=4$ and $n=5$ cases explicitly, keeping in mind the orders of $\alpha\cdots\gamma$ in each configuration.

\vspace*{0.10cm}
\noindent\underline {$n=4$}
\beq
I_{1234}(A,B,C:b)&=&
(D_{12}D_{34}+D_{13}D_{24}+D_{14}D_{23}+D_{12}E_{3}E_{4}+D_{13}E_{2}E_{4}\nonumber\\ &&+D_{14}E_{2}E_{3}+D_{23}E_{1}E_{4}+D_{24}E_{1}E_{3}+D_{34}E_{1}E_{2}\nonumber\\ &&+E_{1}E_{2}E_{3}E_{4})\, I(A,B,C:b)~.
\label{1234}
\eeq
This form is written following the three rules above.
Using the formula, we can calculate the number of configurations: $N_2^4=3$, $N_1^4=6$, $N_0^4=1$.  These numbers agree with the expression of $I_{1234}(A,B,C:b)$ in Eq.\,(\ref{1234}).

\vspace*{0.10cm}
\noindent\underline {$n=5$}
\beq
I_{12345}(A,B,C:b)&=&
(D_{12}D_{34}E_5+D_{12}D_{35}E_{4}+D_{12}D_{45}E_3+D_{13}D_{24}E_5\nonumber\\ &&+D_{13}D_{25}E_{4}+D_{13}D_{45}E_{2}+D_{14}D_{23}E_5+D_{14}D_{25}E_{3}\nonumber\\ &&+D_{14}D_{35}E_{2}+D_{15}D_{23}E_{4}+D_{15}D_{24}E_{3}+D_{15}D_{34}E_{2}\nonumber\\ &&+D_{23}D_{45}E_{1}+D_{24}D_{35}E_{1}+D_{25}D_{34}E_{1}\nonumber\\ &&+D_{12}E_{3}E_{4}E_5+D_{13}E_{2}E_{4}E_5+D_{14}E_{2}E_{3}E_5+D_{15}E_2E_3E_4\nonumber\\ &&+D_{23}E_{1}E_{4}E_5+D_{24}E_{1}E_{3}E_5+D_{25}E_{1}E_{3}E_{4}+D_{34}E_{1}E_{2}E_5\nonumber\\ &&
+D_{35}E_{1}E_{2}E_{4}+D_{45}E_{1}E_{2}E_{3}+E_{1}E_{2}E_{3}E_{4}E_{5})\, I(A,B,C:b)\,.
\eeq
Using the formula, we get: $N_2^5=15$, $N_1^5=10$, $N_0^5=1$.  These numbers agree with the numbers of each configuration in the above expression.  We have derived the above expressions in various ways.  Rules 1, 2, and 3 for the construction of all the partitions have been verified in the mathematical induction method.  With the above rules we are able to write explicitly the derivative formula for any number of momentum integrations and derivatives.

After getting all the terms, we should set $b=0$ and write the integrals in the same notation:
\beq
I_{ixjy\cdots kz}(A,B,C)=I_{ixjy\cdots kz}(A,B,C:b=0)~.
\eeq
In addition, we shall introduce the superscript $(type)$ for the integrals to specify which nucleons interact with each other; that is discussed in Appendix \ref{sec:gauss}. 

\section{Two-body interaction with two tensor correlation operators}\label{sec:two_fd}
We write explicitly the case of the tensor interaction with two tensor correlations:
\beq
\bra \AMD|F_D V F_D|\AMD \ket~,
\eeq
where
\beq
&&F_D V F_D=\fra \sum_{i\ne j}F_{ij}\fra\sum_{k\ne l}V_{kl}\fra\sum_{m\ne n} F_{mn}
=R_2+R_3+R_4+R_5+R_6~.
\eeq
Here, $R_2,\ldots,R_6$ are two- to six-body operators.
We have written the multiple of two two-body interactions (correlations) as the sum of two-, three- and four-body operators:
\beq
VF_D=Q_2+Q_3+Q_4
\eeq
in Eq.\,(\ref{qn}), 
where
\beq
&&Q_2=\fra \sum_{i\ne j}V_{ij}F_{ij}~,\\
&&Q_3=\sum_{i\ne j\ne k}V_{ij}F_{jk}~,\\
&&Q_4=\frac{1}{4} \sum_{i\ne j\ne k \ne l} V_{ij}F_{kl}~.
\eeq
We calculate the multiple of three two-body interactions (correlations) $R_2,\ldots,R_6$ 
baesd on the multiple of two two-body interactions (correlations) $Q_2, Q_3$, and $Q_4$.  
We start with the two-body operator $R_2$ basing on $Q_2$:
\beq
R_2(Q_2)=\left(\fra \sum_{i\ne j} F_{ij} Q_2\right)_{\rm \hspace*{-0.1cm}2\,body}\hspace*{-0.1cm}=\fra \sum_{i\ne j} F_{ij}V_{ij}F_{ij}~.
\eeq
We obtain the three-body operator $R_3$ based on $Q_2$:
\beq
R_3(Q_2)=\left(\fra \sum_{i\ne j} F_{ij} Q_2\right)_{\rm \hspace*{-0.1cm}3\,body}\hspace*{-0.1cm}= \sum_{i\ne j\ne k} F_{ij}V_{jk}F_{jk}~.
\eeq
We have the three-body operator $R_3$ based on $Q_3$:
\beq
R_3(Q_3)=\left(\fra \sum_{i\ne j} F_{ij} Q_3\right)_{\rm \hspace*{-0.1cm}3\,body}\hspace*{-0.1cm}= \sum_{i\ne j\ne k} F_{ij}V_{ij}F_{jk}+\sum_{i\ne j\ne k} F_{jk}V_{ij}F_{jk}+\sum_{i\ne j\ne k} F_{ik}V_{ij}F_{jk}~.
\eeq
We obtain the four-body operator $R_4$ based on $Q_2$:
\beq
R_4(Q_2)=\left(\fra \sum_{i\ne j} F_{ij} Q_2\right)_{\rm \hspace*{-0.1cm}4\,body}\hspace*{-0.1cm}= \frac{1}{4}\sum_{i\ne j\ne k\ne l} F_{ij}V_{kl}F_{kl}~.
\eeq
We obtain the four-body operator $R_4$ based on $Q_3$:
\beq
R_4(Q_3)&=&\left(\fra \sum_{i\ne j} F_{ij} Q_3\right)_{\rm \hspace*{-0.1cm}4\,body}\nonumber\\ &=& \sum_{i\ne j\ne k\ne l} F_{ij}V_{jk}F_{kl}+\sum_{i\ne j\ne k\ne l} F_{ik}V_{jk}F_{kl}+\sum_{i\ne j\ne k\ne l} F_{il}V_{jk}F_{kl}~.
\eeq
We obtain the four-body operator $R_4$ based on $Q_4$:
\beq
R_4(Q_4)&=&\left(\fra \sum_{i\ne j} F_{ij} Q_4\right)_{\rm\hspace*{-0.1cm}4\,body}\nonumber\\ &=& \frac{1}{4}\sum_{i\ne j\ne k\ne l} F_{ij}V_{ij}F_{kl}+\sum_{i\ne j\ne k\ne l} F_{ik}V_{ij}F_{kl}+\frac{1}{4}\sum_{i\ne j\ne k\ne l} F_{kl}V_{ij}F_{kl}~.
\eeq
We obtain the five-body operator $R_5$ based on $Q_3$:
\beq
R_5(Q_3)=\left(\fra \sum_{i\ne j} F_{ij} Q_3\right)_{\rm\hspace*{-0.1cm}5\,body}\hspace*{-0.1cm}= \fra\sum_{i\ne j\ne k\ne l\ne m} F_{ij}V_{kl}F_{lm}~.
\eeq
We obtain the five-body operator $R_5$ based on $Q_4$:
\beq
R_5(Q_4)=\left(\fra \sum_{i\ne j} F_{ij} Q_4\right)_{\rm\hspace*{-0.1cm}5\,body}\hspace*{-0.1cm}= \fra\sum_{i\ne j\ne k\ne l\ne m} F_{ij}V_{jk}F_{lm}+\fra\sum_{i\ne j\ne k\ne l\ne m} F_{il}V_{jk}F_{lm}~.
\eeq
We obtain the six-body operator $R_6$ based on $Q_4$:
\beq
R_6(Q_4)=\left(\fra \sum_{i\ne j} F_{ij} Q_4\right)_{\rm\hspace*{-0.1cm}6\,body}\hspace*{-0.1cm}= \frac{1}{8}\sum_{i\ne j\ne k\ne l\ne m\ne n} F_{ij}V_{kl}F_{mn}~.
\eeq
The symmetry factor in front of the summation is tabulated in Table 1 of Sect. 6.

\subsection{Matrix element of a multiple of three operators}
We start with a two-body operator of the category $type=(12)^3$:
\beq
&&\bra \AMD|R_2(Q_2)|\AMD \ket=\fra \sum_{pqst}\sum_{\mu1\mu2\mu3} \tilde C_{\mu1}^{(2)}\tilde C_{\mu2}^{(2)}\tilde C_{\mu3}^{(2)} \nonumber\\ &&~~
\sum_{xyzuvw\,x'y'z'u'v'w'}I_{1x1y2z2u3v3w}^{((12)^3)}(A,B,C)\nonumber\\ &&~~
(3\delta_{xx'}\delta_{yy'}-\delta_{xy}\delta_{x'y'})(3\delta_{zz'}\delta_{uu'}-\delta_{zu}\delta_{z'u'})(3\delta_{vv'}\delta_{ww'}-\delta_{vw}\delta_{v'w'})\nonumber\\ &&~~M_{x'z'v'}^{ps}M_{y'u'w'}^{qt}\sum_{ab}\bar M_{ab}^{ps}\bar M_{ab}^{qt}C(pq:st)~.
\eeq
The momentum integral $I_{1x1y2z2u3v3w}^{((12)^3)}(A,B,C)$ is given in Sect. \ref{sec:momentum} and Appendix \ref{sec:gauss}, where $(type=(12)^3)$ indicates three two-body operators acting on particles 1 and 2.

We calculate the three-body operator $R_3(Q_2)$ of $type=12:(23)^2$:
\beq
&&\bra \AMD|R_3(Q_2)|\AMD \ket=\sum_{pqrsto}\sum_{\mu1\mu2\mu3} \tilde C_{\mu1}^{(2)}\tilde C_{\mu2}^{(2)}\tilde C_{\mu3}^{(2)}\nonumber\\ &&~~
\sum_{xyzuvw\,x'y'z'u'v'w'}I_{1x1y2z2u3v3w}^{(12:(23)^2)}(A,B,C)\nonumber\\ &&~~
(3\delta_{xx'}\delta_{yy'}-\delta_{xy}\delta_{x'y'})(3\delta_{zz'}\delta_{uu'}-\delta_{zu}\delta_{z'u'})(3\delta_{vv'}\delta_{ww'}-\delta_{vw}\delta_{v'w'})\nonumber\\ &&~~M_{x'}^{ps}M_{y'z'v'}^{qt}M_{u'w'}^{ro}\sum_{ab}\bar M_a^{ps}\bar M_{ab}^{qt}\bar M_b^{ro}C(prq:sto)~.
\eeq

We come to the first term of the three-body operator $R_3(Q_3)$ of the category $type=(12)^2:23$:
\beq
&&\bra \AMD|R_3(Q_3)|\AMD \ket= \sum_{pqrsto}\sum_{\mu1\mu2\mu3} \tilde C_{\mu1}^{(2)}\tilde C_{\mu2}^{(2)}\tilde C_{\mu3}^{(2)}\nonumber\\ &&~~
\sum_{xyzuvw\,x'y'z'u'v'w'}I_{1x1y2z2u3v3w}^{((12)^2:23)}(A,B,C)\nonumber\\ &&~~
(3\delta_{xx'}\delta_{yy'}-\delta_{xy}\delta_{x'y'})(3\delta_{zz'}\delta_{uu'}-\delta_{zu}\delta_{z'u'})(3\delta_{vv'}\delta_{ww'}-\delta_{vw}\delta_{v'w'})\nonumber\\ &&~~M_{x'z'}^{ps}M_{y'u'v'}^{qt}M_{w'}^{ro}\sum_{ab}\bar M_{a}^{ps}\bar M_{ab}^{qt}\bar M_b^{ro}C(prq:sto)~.
\eeq
The other two terms for $R_3(Q_3)$ are categorized as $type=23:12:23$ and $13:12:23$.  The superscripts of the integral $I$ have to be changed according to the type and the spin and isospin matrix elements should be changed slightly for these terms.

We write the four-body operator $R_4(Q_2)$ of the category $type=12:(34)^2$:
\beq
&&\bra \AMD|R_4(Q_2)|\AMD \ket=\frac{1}{4} \sum_{pqrs\,p'q'r's'}\sum_{\mu1\mu2\mu3} \tilde C_{\mu1}^{(2)}\tilde C_{\mu2}^{(2)}\tilde C_{\mu3}^{(2)}\nonumber\\ &&~~
\sum_{xyzuvw\,x'y'z'u'v'w'}I_{1x1y2z2u3v3w}^{(12:(34)^2)}(A,B,C)\nonumber\\ &&~~
(3\delta_{xx'}\delta_{yy'}-\delta_{xy}\delta_{x'y'})(3\delta_{zz'}\delta_{uu'}-\delta_{zu}\delta_{z'u'})(3\delta_{vv'}\delta_{ww'}-\delta_{vw}\delta_{v'w'})\nonumber\\ &&~~M_{x'}^{pp'}M_{y'}^{qq'}M_{z'v'}^{rr'}M_{u'w'}^{ss'}\sum_{ab}\bar M_a^{pp'}\bar M_a^{qq'}\bar M_b^{rr'}\bar M_b^{ss'}C(pqrs:p'q'r's')~.
\eeq
There are many more matrix elements, which are obtained in the same way.  We skip writing these matrix elements here.  Other interactions are written in a similar way by changing the characters of the operators, as discussed in Sect.\,\ref{sec:model}.  The systematic way to calculate matrix elements will be described in Sect.\,\ref{sec:short}.

\section{Short-range correlation}\label{sec:short}

We have to introduce further the short-range correlation in the many-body wave function.  The discussion of the short-range correlation has been delayed up to this section, since the main difficulty with the nuclear many-body problem is the treatment of the tensor correlation.  We have developed various methods to handle the tensor correlation operator.  We shall use the same concept as the tensor correlation for the short-range correlation.

Several methods for the short-range correlation have been developed in the past.  One popular method is the Jastrow correlation operator method~\cite{gaudin71} and another is the unitary correlation operator method (UCOM)~\cite{feldmeier98}.  In the Jastrow method, the correlation operator is written as a product of correlation functions.  In the UCOM, a Hermite correlation operator is placed on an exponential so that the correlation operator is unitary.   In all these methods, the matrix elements are obtained by introducing an approximation to take the resulting operators up to few-body operators.  As discussed for the case of the tensor correlation operator, we are able to add correlation operators systematically one after another to see the convergence of the solutions.  Hence, it is important to know that the present formulation is able to calculate all the matrix elements systematically and straightforwardly.

\subsection{Wave function with short-range correlation}
We introduce the short-range correlation operator $F_S$ in the TOAMD wave function as in the case of the tensor correlation operator.
\beq
|\tilde \Psi\ket=(1+F_S) |\Psi\ket~,
\eeq
where $|\Psi\ket$ was introduced as the TOAMD wave function in Eq.\,(1).  This arrangement indicates that the full wave function is the sum of the following four components:
\beq
|\tilde \Psi \ket=|\AMD\ket + F_S|\AMD\ket + F_D |\AMD\ket + F_SF_D|\AMD \ket~.
\eeq
The first term provides low-momentum components representing the shape of nucleus, the second term provides high-momentum components due to the short-range correlation, and the third term provides intermediate-hight-momentum components due to the tensor correlation.  The last term is an interference term for the short-range and tensor correlations. 
Here, we expand the short-range correlation operator in the sum of Gaussian functions:
\beq
F_S=\fra \sum_{i\ne j}\sum_\mu C_\mu e^{-a_\mu r_{ij}^2}~.
\eeq
These expansion parameters are considered as variational parameters of the many-body wave function.  The short-range correlation is strong in the non-spin, non-isospin channel and we show only this case. However, we will use spin- and isospin-dependent short-range correlations in the calculation.  We can then obtain the many-body Schr\"odinger equation 
$H |\tilde \Psi \ket = E |\tilde \Psi\ket$ and the eigenvalue is:
\beq 
E=\frac{\bra \Psi|(1+F_S) H (1+F_S)| \Psi \ket}{\bra \Psi|(1+F_S) (1+F_S)| \Psi \ket} ~.
\eeq

Although the operator structure of the short-range correlation is simple, there appear many-body operators.  Hence, as an example we discuss the case of the three-body interaction with the short-range and tensor correlations in the next subsection.  We then discuss a general method to calculate all the necessary matrix elements in the subsequent subsection.

\subsection{Three-body interaction with the short-range correlation for the tensor component}
We explicitly write a complicated matrix element where the number of momentum integrals is 6 : 2 from the short-range correlation, 2 from the tensor correlation, and 2 from the three-body interaction.  We consider the case of the repulsive three-body interaction with the short-range correlation:
\beq
F_S U_R F_S 
&=&
\sum_{i\ne j\ne k}\left(\fra A_R\right)\sum_{\mu1 \mu2 \mu3 \mu4} 
C_{\mu1} \left(\frac{\pi}{a_{\mu1}}\right)^{3/2} 
C_{\mu2} \left(\frac{\pi}{a_{\mu2}}\right)^{3/2} 
C_{\mu3} \left(\frac{\pi}{a_{\mu3}}\right)^{3/2} 
\nonumber\\
&&~~C_{\mu4} \left(\frac{\pi}{a_{\mu4}}\right)^{3/2}\ 
\int_{k_1}\int_{k_2}\int_{k_3}\int_{k_4}e^{- k_1^2/4a_{\mu1}} e^{- k_2^2/4a_{\mu2}}e^{- k_3^2/4a_{\mu3}} e^{- k_4^2/4a_{\mu4}}
\nonumber\\
&&~~e^{i\vec k_1(\vec r_i-\vec r_j)}e^{i\vec k_2(\vec r_i-\vec r_j)}e^{i\vec k_3(\vec r_j-\vec r_k)}e^{i\vec k_4(\vec r_i-\vec r_j)} ~.
\label{eq:FUF}
\eeq
Here, we have introduced an approximation that the short-range correlations act only on the same nucleon pairs, $i$ and $j$ in the three-body interaction.  This is because the short-range correlation $F_S$ is large only at very short distances and the probability of more than three nucleons coming to the region of the short-range correlation is negligibly small.
If we multiply the two tensor correlation operators by the three-body operator (\ref{eq:FUF}) of the three-body interaction with two short range correlations $F_SU_RF_S$
, we get many terms:
\beq
F_D F_S U_R F_S F_D=R_3 +R_4+R_5+R_6+R_7~.
\eeq
Here, one of the $R_3$ terms for the $type=(12)^3:23:(12)^2$, where three two-body operators act on particles 1 and 2, one two-body operator acts on particles 2 and 3, and two two-body operators act on particles 1 and 2:
\beq
&&R_3= \sum_{i\ne j\ne k}\sum_{\mu1 \mu2 \mu3 \mu4 \mu5 \mu6} \left(\fra A_R\right)\tilde C_{\mu1}^{(2)}\tilde C_{\mu2}^{(0)}\tilde C_{\mu3}^{(0)}\tilde C_{\mu4}^{(0)}\tilde C_{\mu5}^{(0)}\tilde C_{\mu6}^{(2)} \nonumber\\ &&~~\int_{k_1}\int_{k_2}\int_{k_3}\int_{k_4}\int_{k_5}\int_{k_6} e^{- k_1^2/4a_{\mu1}} e^{- k_2^2/4a_{\mu2}}e^{- k_3^2/4a_{\mu3}} e^{- k_4^2/4a_{\mu4}}e^{- k_5^2/4a_{\mu5}} e^{- k_6^2/4a_{\mu6}}\nonumber\\ &&~~ e^{i\vec k_1(\vec r_i-\vec r_j)}e^{i\vec k_2(\vec r_i-\vec r_j)}e^{i\vec k_3(\vec r_i-\vec r_j)}e^{i\vec k_4(\vec r_j-\vec r_k)}e^{-i\vec k_5 (\vec r_i-\vec r_j)}e^{-i\vec k_6 (\vec r_i-\vec r_j)} \nonumber\\ &&~~\sum_{xyzu\,x'y'z'u'}k_{1x}k_{1y}k_{6z}k_{6u}(3\delta_{xx'}\delta_{yy'}-\delta_{xy}\delta_{x'y'})(3\delta_{zz'}\delta_{uu'}-\delta_{zu}\delta_{z'u'})\nonumber\\ &&~~
\sigma_{ix'}\sigma_{jy'}\sigma_{iz'}\sigma_{ju'}\sum_{vw}\tau_{iv}\tau_{jv}\tau_{iw}\tau_{jw}~.
\eeq
Here we write the case, where two tensor operators and two short-range operators act on $i, j$ pairs and the three-body interaction works for $i, j, k$ nucleons:
\beq
&&\bra \AMD|R_3|\AMD\ket= \sum_{pqr\,p'q'r'}\sum_{\mu1 \mu2 \mu3 \mu4 \mu5 \mu6} \left(\fra A_R\right) \tilde C_{\mu1}^{(2)}\tilde C_{\mu2}^{(0)}\tilde C_{\mu3}^{(0)}\tilde C_{\mu4}^{(0)}\tilde C_{\mu5}^{(0)}\tilde C_{\mu6}^{(2)} \nonumber\\ &&~~ \nonumber\\ &&~~\sum_{xyzu\,x'y'z'u'}I_{1x1y6z6u}^{((12)^3:23:(12)^2)}(A,B,C)(3\delta_{xx'}\delta_{yy'}-\delta_{xy}\delta_{x'y'})(3\delta_{zz'}\delta_{uu'}-\delta_{zu}\delta_{z'u'})\nonumber\\ &&~~M_{x'z'}^{pp'}M_{y'u'}^{qq'}M^{rr'}\sum_{vw}\bar M_{vw}^{pp'}\bar M_{vw}^{qq'}\bar M^{rr'}C(pqr:p'q'r')~.
\eeq
We can further write other terms for the three-body operators in a similar way to that above.  We have in addition the four-, five-, six-, and seven-body operators.  All these matrix elements are written in a similar way to the above expression, with major changes in the momentum integrals and small changes in the spin matrix elements.  Depending on the number of nucleons involved for the multi-body operators, the coefficients of the co-factor matrix change as $C(pq\cdots r:p'q'\cdots r')$.

\subsection{Matrix elements for the general case}
Although we are not able to write all the matrix elements, they are written in a systematic way.  The matrix elements are written for general multi-body operators $O$ with many momentum integrations:
\beq
&&\bra \AMD|O|\AMD\ket=S \sum_{pq\cdots r\,p'q'\cdots r'}\sum_{\mu1 \mu2\cdots \mu n} \tilde C_{\mu1}^{(m1)} \tilde C_{\mu2}^{(m2)}\cdots \tilde C_{\mu n}^{(mn)}  \nonumber\\ &&~~\sum_{xy\cdots z}I_{X(x,y,\ldots ,z)}^{(type)}(A,B,C)F(x,y,\ldots,z)M_{Z(x,y,\ldots,z)}^{pq\cdots r p'q'\cdots r'}\bar M_{U(x,y,\ldots,z)}^{pq\cdots rp'q'\cdots r'}C(pq\cdots r:p'q'\cdots r')~.
\nonumber\\
\eeq
Here, $S$ is a symmetry factor for a many-body operator.  We list $S$ for various configurations up to $n=3$ in Table\,\ref{table}.  The symmetry factors are the same for configurations obtained by the interchange of particle numbers. 
\begin{table}[t]
\centering
\caption{The symmetry factor $S$.}
\begin{tabular}{p{2.2cm}p{2.8cm}p{2.2cm}p{0.5cm}}\hline
 $type$ & $S$ &  $type$ & $S$\\ \hline
 12 & 1/2 &&\\
 $(12)^2$ & 1/2 &  12:23 & 1\\
 12:34 & 1/4 &&\\
 $(12)^3$ & 1/2 &  $(12)^2:23$ & 1\\
 $12:23:13$ & 1 &  $(12)^2:34$ & 1/4\\
 $12:23:14$ & 1 & $12:23:24$ & 1\\
 $12:34:13$ &1 &&\\
 $12:23:45$ & 1/2 &  $12:34:15$ & 1/2\\
 $12:34:56$ & 1/8   & & \\
\hline
\end{tabular}
\label{table}
\end{table}
$pq\cdots rp'q'\cdots r'$ are the quantum numbers of AMD states, where each quantum number $p'$s runs from 1 to $A$.  $\mu1\mu2\cdots \mu n$ are the expansion parameters of interactions and correlations in terms of Gaussian functions and the number of $\mu'$s is $n$ representing the number of momenta.
The expansion factors $\tilde C_{\mu}^{(m)}$ are written as:
\beq
\tilde C_{\mu}^{(m)}=C_\mu \left(\frac{\pi}{a_{\mu}}\right)^{3/2}\left(\frac{-i}{2a_\mu}\right)^m~,
\eeq
where $m=1$ for the spin-orbit interaction, $m=2$ for the tensor, squared angular momentum and squared spin-orbit interactions, and $m=0$ otherwise.  The coordinates $x,y,\ldots,z$ run from $x$ to $z$ of the rectangular coordinates. The momentum integral $I_{X(x,y,\ldots,z)}^{(type)}(A,B,C)$ depends on the type of configurations for the coefficients $A, B, C$ with differentiation of momentum given by a function $X(x,y,\ldots,z)$. $F(x,y,\ldots,z)$ represents the type of interactions and correlations to be specified by the interaction.  $M_{Z(x,y,\ldots,z)}^{pq\cdots rp'q'\cdots r'}$ is the spin matrix element of all the AMD states and $\bar M_{U(x,y,\ldots,z)}^{pq\cdots rp'q'\cdots r'}$ the isospin matrix element.  The co-factor $C(pq\cdots r:p'q'\cdots r')$ takes care of the antisymmetrization of particles in the many-body operators.

We describe the procedure of writing matrix elements for various operators.  We first fix which matrix elements to calculate the configuration: $type$.  Once we fix $type$, we can get the symmetry factor $S$ for this configuration.  We order all the operators from left to right and assign momenta $\vec k_1,\ldots,\vec k_n$.  We then write all the operators explicitly keeping the order of the operators.  For each operator for particles $i$ and $j$, we write the following factors:

\vspace*{0.10cm}
\noindent\underline{Central interaction}\\
We write the case of the spin-spin and isospin-isospin interaction:
\beq
O^c=\sum_{xy} \sigma_{ix}\sigma_{jx}\tau_{iy}\tau_{jy}~,
\eeq
with $m=0$.  If there are no spin- and/or no isospin-dependent operators, we simply drop these spin and isospin operators.

\vspace*{0.10cm}
\noindent\underline{Tensor interaction}\\
\beq
O^t=\sum_{xyx'y'} k_{1x}k_{1y} \sigma_{ix'}\sigma_{jy'} (3\delta_{xx'}\delta_{yy'}-\delta_{xy}\delta_{x'y'})~,
\eeq
with $m=2$.

\vspace*{0.10cm}
\noindent\underline{Spin-orbit interaction}\\
\beq
O^{ls}=\sum_{xyz}\varepsilon_{xyz} k_{1x}\fra(p_i-p_j)_y \fra(\sigma_i+\sigma_j)_z~,
\eeq
with $m=1$.

\vspace*{0.10cm}
\noindent\underline{Squared angular momentum interaction}\\
\beq
O^{l^2}&=&\sum_{xyzuv} \varepsilon_{xyz}\varepsilon_{uvz}  (k_{1x}k_{1u}-2a_\mu\delta_{xu}) \fra(p_i-p_j)_y\fra(p_i-p_j)_v 
 \nonumber\\ && -4a_\mu\sum_{x} k_{1x}\fra(p_{i}-p_{j})_x~,
\eeq
with $m=2$.
For calculation of the matrix elements, we need matrix elements of $p_x=-i\nabla_x$ and $p_x p_y=(-i\nabla_x)(-i \nabla_y)$ for the calculation of angular momenta, which are given in Eqs.\, (\ref{single}) and (\ref{double}).

\vspace*{0.10cm}
\noindent\underline{Squared spin-orbit interaction}\\
\beq
O^{(ls)^2}&=&\sum_{xyzuvw} \varepsilon_{xyz}\varepsilon_{uvw}  \Bigl[(k_{1x}k_{1u}-2a_\mu\delta_{xu}) \fra(p_i-p_j)_y\fra(p_i-p_j)_v 
 \nonumber\\ &&+2a_\mu \delta_{yu}k_{1x}\fra(p_{i}-p_{j})_v\Bigr]\fra(\sigma_{i}+\sigma_{j})_z\fra(\sigma_{i}+\sigma_{j})_w~,
\eeq
with $m=2$.

We can then write the matrix element explicitly.  As an example, we show the case of $type=12:(34)^3:56$ for the tensor operator:
\beq
&&\bra \AMD|F_D F_S V^t F_S F_D|\AMD\ket\nonumber\\ &&~~=\frac{1}{8}\sum_{pqrsto\,p'q'r's't'o'}\sum_{\mu1\mu2\mu3\mu4\mu5}\tilde C_{\mu1}^{(2)}\tilde C_{\mu2}^{(0)}\tilde C_{\mu3}^{(2)}\tilde C_{\mu4}^{(0)}\tilde C_{\mu5}^{(2)}\nonumber\\ &&~~\sum_{xyzuvw\,x'y'z'u'v'w'}I_{1x1y3z3u5v5w}^{(12:(34)^3:56)}(A,B,C)\nonumber\\ &&~~(3\delta_{xx'}\delta_{yy'}-\delta_{xy}\delta_{x'y'})(3\delta_{zz'}\delta_{uu'}-\delta_{zu}\delta_{z'u'})(3\delta_{vv'}\delta_{ww'}-\delta_{vw}\delta_{v'w'})\nonumber\\ &&~~M_{x'}^{pp'}M_{y'}^{qq'}M_{z'}^{rr'}M_{u'}^{ss'}M_{v'}^{tt'}M_{w'}^{oo'}\nonumber\\ &&~~\sum_{ab}\bar M_a^{pp'}\bar M_a^{qq'}\bar M^{rr'}\bar M^{ss'}\bar M_b^{tt'}\bar M_b^{oo'}C(pqrsto:p'q'r's't'o')~.
\eeq
The momentum integral is given in Appendix \ref{sec:gauss}.  All matrix elements can be written in a similar systematic manner.

As for the spin-orbit interaction with the short-range and tensor correlations, we give explicitly the case of $type=12:(34)^3:56$.
\beq
&&\bra \AMD|F_D F_S V^{ls} F_S F_D|\AMD\ket\nonumber\\ &&~~=\frac{1}{8}\sum_{pqrsto\,p'q'r's't'o'}\sum_{\mu1\mu2\mu3\mu4\mu5}\tilde C_{\mu1}^{(2)}\tilde C_{\mu2}^{(0)}\tilde C_{\mu3}^{(1)}\tilde C_{\mu4}^{(0)}\tilde C_{\mu5}^{(2)}\nonumber\\ &&~~\sum_{xyx'y'zuavwv'w'}\fra \Bigl[I_{1x1y3z4u5v5w}^{(12:(34)^3:56)}(A,B,C)- I_{1x1y3z2u5v5w}^{(12:(34)^3:56)}(A,B,C)\nonumber\\ &&~~+ I_{1x1y3z5v5w}^{(12:(34)^3:56)}(A,B,C)i\nu((D_r-D_{r'})-(D_s-D_{s'}))_u\Bigr]\nonumber\\ &&~~(3\delta_{xx'}\delta_{yy'}-\delta_{xy}\delta_{x'y'})\varepsilon_{zua} (3\delta_{vv'}\delta_{ww'}-\delta_{vw}\delta_{v'w'})\nonumber\\ &&~~M_{x'}^{pp'}M_{y'}^{qq'}\fra(M_{a}^{rr'}M^{ss'}+M^{rr'}M_{a}^{ss'})M_{v'}^{tt'}M_{w'}^{oo'}\nonumber\\ &&~~\sum_{bc}\bar M_b^{pp'}\bar M_b^{qq'}\bar M^{rr'}\bar M^{ss'}\bar M_c^{tt'}\bar M_c^{oo'}C(pqrsto:p'q'r's't'o')~.
\eeq

The spin-orbit interaction has a derivative term.  Therefore the expressions are slightly different for different configurations.  We write one similar configuration case: $type=12:(34)^3:35$.
\beq
&&\bra \AMD|F_D F_S V^{ls} F_S F_D|\AMD\ket\nonumber\\ &&~~=\frac{1}{2}\sum_{pqrst\,p'q'r's't'}\sum_{\mu1\mu2\mu3\mu4\mu5}\tilde C_{\mu1}^{(2)}\tilde C_{\mu2}^{(0)}\tilde C_{\mu3}^{(1)}\tilde C_{\mu4}^{(0)}\tilde C_{\mu5}^{(2)}\nonumber\\ &&~~\sum_{xyx'y'zuavwv'w'}\fra\Bigl[I_{1x1y3z4u5v5w}^{(12:(34)^3:35)}(A,B,C)+\fra I_{1x1y3z5u5v5w}^{(12:(34)^3:35)}(A,B,C)\nonumber\\ &&~~- I_{1x1y3z2u5v5w}^{(12:(34)^3:35)}(A,B,C)+ I_{1x1y3z5v5w}^{(12:(34)^3:35)}(A,B,C)i\nu((D_r-D_{r'})-(D_s-D_{s'}))_u\Bigr]\nonumber\\ &&~~(3\delta_{xx'}\delta_{yy'}-\delta_{xy}\delta_{x'y'})\varepsilon_{zua} (3\delta_{vv'}\delta_{ww'}-\delta_{vw}\delta_{v'w'})M_{x'}^{pp'}M_{y'}^{qq'}M_{w'}^{tt'}\fra(M_{av'}^{rr'}M^{ss'}+M_{v'}^{rr'}M_{a}^{ss'})\nonumber\\ &&~~\sum_{bc}\bar M_b^{pp'}\bar M_b^{qq'}\bar M_c^{rr'}\bar M^{ss'}\bar M_c^{tt'}C(pqrst:p'q'r's't')~.
\eeq

\section{Summary}\label{sec:summary}

We have developed a powerful many-body theory to describe finite nuclei, which is calld ``tensor-optimized antisymmetrized molecular dynamics'' (TOAMD).  The TOAMD theory is based on AMD, in which the concept of the TOSM is incorporated, in order to treat the strong tensor interaction in the nucleon-nucleon interaction.  The tensor interaction is treated by the tensor correlation operator acting on the AMD wave function.  Since the tensor interaction is of long and intermediate range, we have to explicitly treat many-body operators due to the tensor correlation operators and the two- and three-body interactions in the Hamiltonian for the AMD state.  For the TOAMD theory, we have to treat up to 6-body operator terms for the two-body interaction and 7-body operator terms for the three-body interaction for the AMD state.

In order to treat multi-body operators for many-body nuclear systems efficiently, we should use all the powerful mathematics.  The AMD wave function consists of shifted Gaussian functions with spin and isospin wave functions and all the interactions are expanded in Gaussian functions.  In addition, we have to take into account the antisymmetrization of all the nucleons.  For this purpose, we take the Fourier transforms of all the interactions so that any multi-body operators are written in separable forms in particle coordinates.  We are then able to calculate multi-body matrix elements with antisymmetrization using the multi body co-factor matrix of the norm matrix.  We should then take the multi-momentum integrations, where we have developed a systematic integral and differentiation method with source terms.  The final results are written as the sum of many terms for the norm and energy matrices, which do not involve any numerical integrations.  We have to minimize the total energy with respect to the variational parameters, which are the shift coordinates and spin weights in the AMD wave function, the tensor correlation function, and additionally, the short-range correlation function.

The TOAMD theory provides the total energy as function of the variational parameters, where the total energy can be calculated systematically in a straightforward manner.  Since the matrix elements can be calculated in systematic methods, we are able to improve the calculated results by adding more and more complicated correlations including both the tensor and short-range correlations. For heavy nuclei, we have to perform the angular momentum projection and take the sum of the Slater determinants for better description of many-body systems.  The TOAMD theory is a powerful and economical method to treat nuclear many-body system.  The formulation is transparent and we are able to calculate two- and three-body interactions with any order of tensor and short-range correlations. In the TOAMD theory, we should be able to calculate nuclei with many nucleons using the present powerful computers.

In the very near future, we shall publish numerical results of s-shell nuclei, and successively, p-shell and heavier nuclei in the TOAMD theory using the bare nucleon-nucleon interaction. 

\ack
We are grateful to Prof. Akihiro Tohsaki for fruitful discussions and further developments.  We appreciate the hospitality of the RCNP theory group.  We acknowledge the support of JSPS: 23224008, 24740175, 25400256, 25887049, and 15K05091.

\appendix
\section{Property of a matrix for antisymmetrization}\label{sec:matrix}
We give here the necessary formula to calculate multi-body operators, which are separable in particle coordinates: $M(i,j,\ldots,k)=O(i)O(j)\cdots O(k)$. We start with the one-body operator and then given the matrix elements of multi-body operators.

\vspace*{0.10cm}
\noindent\underline{\it One-body operator}\\
We should calculate the matrix element of the one-body operator:
\beq
&&\bra p_1 p_2 \cdots p_A| \sum_{i=1}^A O(i) |\det|q_1 q_2 \cdots q_A| \ket=\sum_{r=1}^A  \left | \begin{array}{ccccc}
\bra p_1|q_1 \ket & \bra p_1|q_2 \ket & \cdots & \bra p_1|q_A\ket\\
\vdots & \vdots & \ddots & \vdots \\
\bra p_r|O|q_1 \ket & \bra p_r|O|q_2 \ket & \cdots & \bra p_r|O|q_A\ket\\
\vdots & \vdots & \ddots & \vdots \\
\bra p_A|q_1 \ket & \bra p_A|q_2 \ket & \cdots & \bra p_A|q_A\ket\\
\end{array}\right |\nonumber\\ && =\sum_{r=1}^A\sum_{l=1}^A \bra p_r|O|q_l \ket\ C(r:l)~.
\label{oneb}
\eeq
Here, $C(r:l)$ is the determinant of a co-factor matrix of $B$, where the $r$-row and  $l$-column are removed from the $A\times A$ matrix.  From this construction, $C(r:l)$ is a function of the single-particle overlap $\bra p_i|q_j \ket$ and does not depend on the one-body matrix element.  We shall derive $C$ for the general case later.

\vspace*{0.10cm}
\noindent\underline{\it Two-body operator}
\beq
\nonumber
&&\bra p_1 p_2\cdots p_A| \sum_{i\ne j}^A O(i)O(j) |\det|q_1 q_2\cdots q_A| \ket=\sum_{r_1\ne r_2}^A  \left | \begin{array}{ccccc}
\bra p_1|q_1 \ket & \bra p_1|q_2 \ket & \cdots & \bra p_1|q_A\ket\\
\vdots & \vdots & \ddots & \vdots \\
\bra p_{r_1}|O|q_1 \ket & \bra p_{r_1}|O|q_2 \ket & \cdots & \bra p_{r_1}|O|q_A\ket\\
\vdots & \vdots & \ddots & \vdots \\
\bra p_{r_2}|O|q_1 \ket & \bra p_{r_2}|O|q_2 \ket & \cdots & \bra p_{r_2}|O|q_A\ket\\
\vdots & \vdots & \ddots & \vdots \\
\bra p_A|q_1 \ket & \bra p_A|q_2 \ket & \cdots & \bra p_A|q_A\ket\\
\end{array}\right |\\&& =\sum_{r_1\ne r_2,l_1\ne l_2}^A \bra p_{r_1}|O|q_{l_1} \ket \bra p_{r_2}|O|q_{l_2} \ket\ C({r_1}{r_2}:{l_1}{l_2})~.
\label{twob}
\eeq
Here, $C(r_1 r_2: l_1 l_2)$ is the determinant of a co-factor matrix of $B$, where the ${r_1}$ and ${r_2}$ rows and  ${l_1}$ and ${l_2}$ columns are removed from the $A\times A$ matrix.  Again $C(r_1 r_2: l_1 l_2)$ is a function of the single particle overlap $\bra p_i|q_j \ket$ and does not depend on the one-body matrix element.

\vspace*{0.10cm}
\noindent\underline{\it Multi-body operator}
\beq
&&\bra p_1 p_2\cdots{p_A}| \sum_{i\ne j\ne\cdots\ne k}^A O(i)O(j)\cdots O(k) |\det|q_1 q_2\cdots q_A| \ket\nonumber\\ && 
=\sum_{r_1\ne r_2\ne \cdots{r_k},l_1\ne l_2\ne\cdots{l_k}}^A \hspace*{-0.5cm} 
\bra p_{r_1}|O|q_{l_1} \ket \bra p_{r_2}|O|q_{l_2} \ket \cdots \bra p_{r_k}|O|q_{l_k} \ket\ C(r_1 r_2 \cdots{r_k}:l_1 l_2\cdots{l_k})~.
\label{multb}
\eeq
Here, $C(r_1 r_2\cdots r_k: l_1 l_2\cdots r_k)$ is the determinant of a co-factor matrix of $B$, where the ${r_1},{r_2},\ldots,r_k$ rows and  ${l_1},{l_2},\ldots,l_k$ columns are removed from the $A\times A$ matrix. Again, $C(r_1 r_2 \cdots r_k: l_1 l_2\cdots l_k)$ is a function of the single-particle overlap $\bra p_i|q_j \ket$ and does not depend on the one-body matrix elements.

\vspace*{0.10cm}
\noindent\underline{\it Determinant of the co-factor matrix}\\
We derive here the determinant $C({r_1}{r_2}\cdots{r_k}:{l_1}{l_2}\cdots{l_k})$ of the co-factor matrix of the $A\times A$ matrix $B$.  We have the following identity for the $A\times A$ matrix $B$ with the matrix elements $a_{ij}$:
\beq
1=\sum_{l_1\ne l_2\ne\cdots \ne l_r}^A a_{k_1l_1}a_{k_2l_2}\cdots a_{k_rl_r}\left |
\begin{array}{cccc}
(B^{-1})_{l_1k_1} & (B^{-1})_{l_1k_2} & \cdots & (B^{-1})_{l_1k_r}\\
(B^{-1})_{l_2k_1} & (B^{-1})_{l_2k_2} & \cdots & (B^{-1})_{l_2k_r}\\
\vdots & \vdots & \ddots & \vdots \\
(B^{-1})_{l_rk_1} & (B^{-1})_{l_rk_2} & \cdots & (B^{-1})_{l_rk_r}\\
\end{array}\right  |
\eeq
Here, $B^{-1}$ is the inverse matrix of $B$.  Since this is an important formula to get the determinant of the co-factor matrix, we verify this explicitly using the definition of the determinant:
\beq
&&({\rm rhs})=\sum_{l_1\ne l_2\ne\cdots\ne l_r}^A a_{k_1l_1}a_{k_2l_2}\cdots a_{k_rl_r}\sum_{P}\epsilon(P) (B^{-1})_{l_1 k_{P(1)}}(B^{-1})_{l_2 k_{P(2)}}\cdots (B^{-1})_{l_r k_{P(r)}}\nonumber\\ &&
=\sum_{P}\epsilon(P)\sum_{l_1}a_{k_1 l_1}(B^{-1})_{l_1 k_{P(1)}}\sum_{l_2}a_{k_2 l_2}(B^{-1})_{l_2 k_{P(2)}}\cdots\sum_{l_r} a_{k_r l_r}(B^{-1})_{l_r k_{P(r)}}\nonumber\\ &&
=\sum_P \epsilon(P) \delta_{1P(1)}\delta_{2P(2)}\cdots\delta_{rP(r)}=1~.
\eeq
Hence, we can write the determinant of $B$ as:
\beq
\nonumber
&&\left |
\begin{array}{cccc}
a_{11} & a_{12} & \cdots & a_{1A}\\
a_{21} & a_{22} & \cdots & a_{2A}\\
\vdots & \vdots & \ddots & \vdots \\
a_{A1} & a_{A2} & \cdots & a_{AA}\\
\end{array}\right  |= \sum_{l_1\ne l_2\ne\cdots\ne l_r}^A \hspace*{-0.3cm} a_{k_1l_1}a_{k_2l_2}\cdots a_{k_rl_r} \left |
\begin{array}{cccc}
(B^{-1})_{l_1k_1} & (B^{-1})_{l_1k_2} & \cdots & (B^{-1})_{l_1k_r}\\
(B^{-1})_{l_2k_1} & (B^{-1})_{l_2k_2} & \cdots & (B^{-1})_{l_2k_r}\\
           \vdots &            \vdots & \ddots & \vdots \\
(B^{-1})_{l_rk_1} & (B^{-1})_{l_rk_2} & \cdots & (B^{-1})_{l_rk_r}\\
\end{array}\right | \\&& \times  \det|B|
=\sum_{l_1\ne l_2\ne\cdots \ne l_r}^A a_{k_1l_1}a_{k_2l_2}\cdots a_{k_rl_r}C(k_1\cdots k_r:l_1\cdots l_r)~.
\eeq
In the above formula, $k_1, k_2,\ldots,k_r$ are any row numbers of the original matrix $B$.
Here, $C(k_1\cdots k_r:l_1\cdots l_r)$ is the determinant of a co-factor matrix of $B$, where $k_1$, $k_2$,$\ldots$ and $k_r$ rows and $l_1$, $l_2$,$\ldots$ and $l_r$ columns are removed from the matrix $B$.  The coefficient $C(k_1\cdots k_r:l_1\cdots l_r)$ is written using $a_{ij}$ of the original matrix $B$.    Explicitly, the coefficient $C$ is given as
\beq
C(k_1\cdots k_r:l_1\cdots l_r)=\left |
\begin{array}{cccc}
(B^{-1})_{l_1k_1} & (B^{-1})_{l_1k_2} & \cdots & (B^{-1})_{l_1k_r}\\
(B^{-1})_{l_2k_1} & (B^{-1})_{l_2k_2} & \cdots & (B^{-1})_{l_2k_r}\\
\vdots            & \vdots            & \ddots & \vdots           \\
(B^{-1})_{l_rk_1} & (B^{-1})_{l_rk_2} & \cdots & (B^{-1})_{l_rk_r}\\
\end{array}\right |   \det|B|~.
\label{genb}
\eeq
Compared with the matrices of one-, two- and multi-body operators ((\ref{oneb}), (\ref{twob}), and (\ref{multb})), the coefficients $C(k_1\cdots k_r:l_1\cdots l_r)$ are the same as that given in Eq.\,(\ref{genb}).  In the case $r=A$, the co-factor $C$ becomes $\epsilon(P(k_1\cdots k_r:l_1\cdots l_r))$, which is the phase of indicated permutation.  Hence, for $r=A$, the rhs of Eq.\,(\ref{multb}) is simply the determinant of the full matrix of $\bra p_i|O| q_j \ket$.

\section{Gaussian integrals}\label{sec:gauss}
We give here all the necessary Gaussian integrals:
\beq
I^{(type)}(A,B,C)&=&\int_{k_1}\int_{k_2}\cdots\int_{k_n} \bra f_p|e^{iK_1(k_1,\cdots,k_n)r}| g_{p'}\ket\cdots\bra f_q|e^{iK_m(k_1,\cdots,k_n)r}|g_{q'}\ket\nonumber\\ 
&=&
\frac{1}{(2\pi)^{3n}}\left(\frac{\pi^n}{\det |A|}\right)^{3/2}e^{-B^\dagger A^{-1}B/4+C}~.
\eeq
Here, $type$ denotes which operators act between which particles.  The functions $K_i$ are functions of $k_i$, whose explicit forms depend on the type of multi-body operators.   In order to understand the meaning of $A, B, C$, we write the single-particle matrix element:
\beq
\bra \psi_p|e^{ikr}|\psi_q\ket=e^{-\fra \nu(D_p-D_q)^2+\fra i \vec k (\vec D_p+\vec D_q)-\frac{k^2}{8\nu}}~.
\eeq 
We write here all the possible integrals up to the three two-body operators.  There are still several Gaussian integrals necessary for calculations of matrix elements, but they can be obtained in a similar way to those presented in this appendix. 

\vspace*{0.1cm}
\noindent\underline{$type=12$}
\beq
A&=&1/4\nu+1/4a_\mu \nonumber\\ 
B&=&\fra(D_p+D_{p'})-\fra(D_q+D_{q'})\nonumber\\ 
C&=&-\fra \nu[(D_p-D_{p'})^2+(D_q-D_{q'})^2]
\eeq
In this appendix, we write vectors $(\vec D_p)'$s simply without the vector notation as $D_p$.  We write here $type=12$, which indicates that a two-body operator act on particles 1 and 2.

\vspace*{0.1cm}
\noindent\underline{$type=(12)^2$}
\beq
A&=&\left(\begin{array}{cc}
1/4\nu+1/4a_{\mu1}&1/4\nu\\ 
1/4\nu & 1/4\nu+1/4a_{\mu2}\\
\end{array}\right) \nonumber\\ 
B&=&\left(\begin{array}{c}
\fra(D_p+D_{p'})-\fra(D_q+D_{q'})\\
\fra(D_p+D_{p'})-\fra(D_q+D_{q'})\\
\end{array}\right)\nonumber\\ 
C&=&-\fra \nu[(D_p-D_{p'})^2+(D_q-D_{q'})^2]
\eeq
Here $type=(12)^2$ means that two two-body operators act on particles 1 and 2.

\vspace*{0.1cm}
\noindent\underline{$type=12:23$}
\beq
A&=&\left(\begin{array}{cc}
1/4\nu+1/4a_{\mu1}&-1/8\nu\\ 
-1/8\nu & 1/4\nu+1/4a_{\mu2}\\
\end{array}\right) \nonumber\\ 
B&=&\left(\begin{array}{c}
\fra(D_p+D_{p'})-\fra(D_q+D_{q'})\\
\fra(D_q+D_{q'})-\fra(D_r+D_{r'})\\
\end{array}\right)\nonumber\\ 
C&=&-\fra \nu[(D_p-D_{p'})^2+(D_q-D_{q'})^2+(D_r-D_{r'})^2]
\eeq
Here, $type=12:23$ means that one interaction acts on particles 1 and 2 and the other on particles 2 and 3. We omit the explanation of $type$ in the following.

\vspace*{0.1cm}
\noindent\underline{$type=12:34$}
\beq
A&=&\left(\begin{array}{cc}
1/4\nu+1/4a_{\mu1}&0\\ 
0 & 1/4\nu+1/4a_{\mu2}\\
\end{array}\right) \nonumber\\ 
B&=&\left(\begin{array}{c}
\fra(D_p+D_{p'})-\fra(D_q+D_{q'})\\
\fra(D_r+D_{r'})-\fra(D_s+D_{s'})\\
\end{array}\right)\nonumber\\ 
C&=&-\fra\nu[(D_p-D_{p'})^2+(D_q-D_{q'})^2+(D_r-D_{r'})^2+(D_s-D_{s'})^2]
\eeq

\vspace*{0.1cm}
\noindent\underline{$type=(12)^3$}
\beq
A&=&\left(\begin{array}{ccc}
1/4\nu+1/4a_{\mu1}&1/4\nu & 1/4\nu\\ 
1/4\nu & 1/4\nu+1/4a_{\mu2} & 1/4\nu\\
1/4\nu & 1/4\nu & 1/4\nu+1/4a_{\mu3}\\
\end{array}\right) \nonumber\\ 
B&=&\left(\begin{array}{c}
\fra(D_p+D_{p'})-\fra(D_q+D_{q'})\\
\fra(D_p+D_{p'})-\fra(D_q+D_{q'})\\
\fra(D_p+D_{p'})-\fra(D_q+D_{q'})\\
\end{array}\right)\nonumber\\ 
C&=&-\fra \nu[(D_p-D_{p'})^2+(D_q-D_{q'})^2]
\eeq

\vspace*{0.1cm}
\noindent\underline{$type=(12)^2:23$}
\beq
A&=&\left(\begin{array}{ccc}
1/4\nu+1/4a_{\mu1}&1/4\nu & -1/8\nu\\ 
1/4\nu & 1/4\nu+1/4a_{\mu2} & -1/8\nu\\
-1/8\nu & -1/8\nu & 1/4\nu+1/4a_{\mu3}\\
\end{array}\right) \nonumber\\ 
B&=&\left(\begin{array}{c}
\fra(D_p+D_{p'})-\fra(D_q+D_{q'})\\
\fra(D_p+D_{p'})-\fra(D_q+D_{q'})\\
\fra(D_q+D_{q'})-\fra(D_r+D_{r'})\\
\end{array}\right)\nonumber\\ 
C&=&-\fra\nu[(D_p-D_{p'})^2+(D_q-D_{q'})^2+(D_r-D_{r'})^2]
\eeq

\vspace*{0.1cm}
\noindent\underline{$type=12:(23)^2$}
\beq
A&=&\left(\begin{array}{ccc}
1/4\nu+1/4a_{\mu1}&-1/8\nu & -1/8\nu\\ 
-1/8\nu & 1/4\nu+1/4a_{\mu2} & 1/4\nu\\
-1/8\nu & 1/4\nu & 1/4\nu+1/4a_{\mu3}\\
\end{array}\right) \nonumber\\ 
B&=&\left(\begin{array}{c}
\fra(D_p+D_{p'})-\fra(D_q+D_{q'})\\
\fra(D_q+D_{q'})-\fra(D_r+D_{r'})\\
\fra(D_q+D_{q'})-\fra(D_r+D_{r'})\\
\end{array}\right)\nonumber\\ 
C&=&-\fra\nu[(D_p-D_{p'})^2+(D_q-D_{q'})^2+(D_r-D_{r'})^2]
\eeq

\vspace*{0.1cm}
\noindent\underline{$type=12:23:13$}
\beq
A&=&\left(\begin{array}{ccc}
1/4\nu+1/4a_{\mu1}&-1/8\nu & 1/8\nu\\ 
-1/8\nu & 1/4\nu+1/4a_{\mu2} & 1/8\nu\\
1/8\nu & 1/8\nu & 1/4\nu+1/4a_{\mu3}\\
\end{array}\right) \nonumber\\ 
B&=&\left(\begin{array}{c}
\fra(D_p+D_{p'})-\fra(D_q+D_{q'})\\
\fra(D_q+D_{q'})-\fra(D_r+D_{r'})\\
\fra(D_p+D_{p'})-\fra(D_r+D_{r'})\\
\end{array}\right)\nonumber\\ 
C&=&-\fra\nu[(D_p-D_{p'})^2+(D_q-D_{q'})^2+(D_r-D_{r'})^2]
\eeq

\vspace*{0.1cm}
\noindent\underline{$type=(12)^2:34$}
\beq
A&=&\left(\begin{array}{ccc}
1/4\nu+1/4a_{\mu1}&1/4\nu & 0\\ 
1/4\nu & 1/4\nu+1/4a_{\mu2} & 0\\
0 & 0 & 1/4\nu+1/4a_{\mu3}\\
\end{array}\right) \nonumber\\ 
B&=&\left(\begin{array}{c}
\fra(D_p+D_{p'})-\fra(D_q+D_{q'})\\
\fra(D_p+D_{p'})-\fra(D_q+D_{q'})\\
\fra(D_r+D_{r'})-\fra(D_s+D_{s'})\\
\end{array}\right)\nonumber\\ 
C&=&-\fra\nu[(D_p-D_{p'})^2+(D_q-D_{q'})^2+(D_r-D_{r'})^2+(D_s-D_{s'})^2]
\eeq

\vspace*{0.1cm}
\noindent\underline{$type=12:23:14$}
\beq
A&=&\left(\begin{array}{ccc}
1/4\nu+1/4a_{\mu1}&-1/8\nu & 1/8\nu\\ 
-1/8\nu & 1/4\nu+1/4a_{\mu2} & 0\\
1/8\nu & 0 & 1/4\nu+1/4a_{\mu3}\\
\end{array}\right) \nonumber\\ 
B&=&\left(\begin{array}{c}
\fra(D_p+D_{p'})-\fra(D_q+D_{q'})\\
\fra(D_q+D_{q'})-\fra(D_r+D_{r'})\\
\fra(D_p+D_{p'})-\fra(D_s+D_{s'})\\
\end{array}\right)\nonumber\\ 
C&=&-\fra\nu[(D_p-D_{p'})^2+(D_q-D_{q'})^2+(D_r-D_{r'})^2+(D_s-D_{s'})^2]
\eeq

\vspace*{0.1cm}
\noindent\underline{$type=12:23:24$}
\beq
A&=&\left(\begin{array}{ccc}
1/4\nu+1/4a_{\mu1}&-1/8\nu & -1/8\nu\\ 
-1/8\nu & 1/4\nu+1/4a_{\mu2} & 1/8\nu\\
-1/8\nu & 1/8\nu & 1/4\nu+1/4a_{\mu3}\\
\end{array}\right) \nonumber\\ 
B&=&\left(\begin{array}{c}
\fra(D_p+D_{p'})-\fra(D_q+D_{q'})\\
\fra(D_q+D_{q'})-\fra(D_r+D_{r'})\\
\fra(D_q+D_{q'})-\fra(D_s+D_{s'})\\
\end{array}\right)\nonumber\\ 
C&=&-\fra\nu[(D_p-D_{p'})^2+(D_q-D_{q'})^2+(D_r-D_{r'})^2+(D_s-D_{s'})^2]
\eeq

\vspace*{0.1cm}
\noindent\underline{$type=12:23:34$}
\beq
A&=&\left(\begin{array}{ccc}
1/4\nu+1/4a_{\mu1}&-1/8\nu & 0\\ 
-1/8\nu & 1/4\nu+1/4a_{\mu2} & -1/8\nu\\
0 & -1/8\nu & 1/4\nu+1/4a_{\mu3}\\
\end{array}\right) \nonumber\\ 
B&=&\left(\begin{array}{c}
\fra(D_p+D_{p'})-\fra(D_q+D_{q'})\\
\fra(D_q+D_{q'})-\fra(D_r+D_{r'})\\
\fra(D_r+D_{r'})-\fra(D_s+D_{s'})\\
\end{array}\right)\nonumber\\ 
C&=&-\fra\nu[(D_p-D_{p'})^2+(D_q-D_{q'})^2+(D_r-D_{r'})^2+(D_s-D_{s'})^2]
\eeq

\vspace*{0.1cm}
\noindent\underline{$type=12:23:45$}
\beq
A&=&\left(\begin{array}{ccc}
1/4\nu+1/4a_{\mu1}&-1/8\nu & 0\\ 
-1/8\nu & 1/4\nu+1/4a_{\mu2} & 0\\
0 & 0 & 1/4\nu+1/4a_{\mu3}\\
\end{array}\right) \nonumber\\ 
B&=&\left(\begin{array}{c}
\fra(D_p+D_{p'})-\fra(D_q+D_{q'})\\
\fra(D_q+D_{q'})-\fra(D_r+D_{r'})\\
\fra(D_s+D_{s'})-\fra(D_t+D_{t'})\\
\end{array}\right)\\ 
C&=&-\fra\nu[(D_p-D_{p'})^2+(D_q-D_{q'})^2+(D_r-D_{r'})^2+(D_s-D_{s'})^2+(D_t-D_{t'})^2]
\nonumber
\eeq

\vspace*{0.1cm}
\noindent\underline{$type=12:34:15$}
\beq
A&=&\left(\begin{array}{ccc}
1/4\nu+1/4a_{\mu1}&0 & 1/8\nu\\ 
0 & 1/4\nu+1/4a_{\mu2} & 0\\
1/8\nu & 0 & 1/4\nu+1/4a_{\mu3}\\
\end{array}\right) \nonumber\\ 
B&=&\left(\begin{array}{c}
\fra(D_p+D_{p'})-\fra(D_q+D_{q'})\\
\fra(D_r+D_{r'})-\fra(D_s+D_{s'})\\
\fra(D_p+D_{p'})-\fra(D_t+D_{t'})\\
\end{array}\right)\\ 
C&=&-\fra\nu[(D_p-D_{p'})^2+(D_q-D_{q'})^2+(D_r-D_{r'})^2+(D_s-D_{s'})^2+(D_t-D_{t'})^2]
\nonumber
\eeq

\vspace*{0.1cm}
\noindent\underline{$type=12:34:35$}
\beq
A&=&\left(\begin{array}{ccc}
1/4\nu+1/4a_{\mu1}&0 & 0\\ 
0 & 1/4\nu+1/4a_{\mu2} & 1/8\nu\\
0 & 1/8\nu & 1/4\nu+1/4a_{\mu3}\\
\end{array}\right) \nonumber\\ 
B&=&\left(\begin{array}{c}
\fra(D_p+D_{p'})-\fra(D_q+D_{q'})\\
\fra(D_r+D_{r'})-\fra(D_s+D_{s'})\\
\fra(D_r+D_{r'})-\fra(D_t+D_{t'})\\
\end{array}\right)\\ 
C&=&-\fra\nu[(D_p-D_{p'})^2+(D_q-D_{q'})^2+(D_r-D_{r'})^2+(D_s-D_{s'})^2+(D_t-D_{t'})^2]
\nonumber
\eeq

\vspace*{0.1cm}
\noindent\underline{$type=12:34:56$}
\beq
A&=&\left(\begin{array}{ccc}
1/4\nu+1/4a_{\mu1}&0 & 0\\ 
0 & 1/4\nu+1/4a_{\mu2} & 0\\
0 & 0 & 1/4\nu+1/4a_{\mu3}\\
\end{array}\right) \nonumber\\ 
B&=&\left(\begin{array}{c}
\fra(D_p+D_{p'})-\fra(D_q+D_{q'})\\
\fra(D_r+D_{r'})-\fra(D_s+D_{s'})\\
\fra(D_t+D_{t'})-\fra(D_u+D_{u'})\\
\end{array}\right)\\ 
C&=&-\fra\nu[(D_p-D_{p'})^2+(D_q-D_{q'})^2+(D_r-D_{r'})^2+(D_s-D_{s'})^2+(D_t-D_{t'})^2+(D_u-D_{u'})^2]
\nonumber
\eeq

We give a more complicated case as an example so that the general rule for $A, B, C$ can be understood instead of writing all the cases.  In this example, $type=12:(34)^3:56$ indicates that one two-body operator acts on particles 1 and 2, three two-body operators on particles 3 and 4, and one two-body operator on particles 5 and 6.

\vspace*{0.1cm}
\noindent\underline{$type=12:(34)^3:56$}
\beq
A&=&\left(\begin{array}{ccccc}
1/4\nu+1/4a_{\mu1}&0 & 0 & 0 & 0\\ 
0 & 1/4\nu+1/4a_{\mu2} & 1/4\nu & 1/4\nu & 0\\
0 & 1/4\nu & 1/4\nu+1/4a_{\mu3} & 1/4\nu & 0\\
0 & 1/4\nu & 1/4\nu & 1/4\nu+1/4a_{\mu4} & 0\\
0 & 0 & 0 & 0& 1/4\nu+1/4a_{\mu5}\\
\end{array}\right) \nonumber\\ 
B&=&\left(\begin{array}{c}
\fra(D_p+D_{p'})-\fra(D_q+D_{q'})\\
\fra(D_r+D_{r'})-\fra(D_s+D_{s'})\\
\fra(D_r+D_{r'})-\fra(D_s+D_{s'})\\
\fra(D_r+D_{r'})-\fra(D_s+D_{s'})\\
\fra(D_t+D_{t'})-\fra(D_u+D_{u'})\\
\end{array}\right)\\ 
C&=&-\fra\nu[(D_p-D_{p'})^2+(D_q-D_{q'})^2+(D_r-D_{r'})^2+(D_s-D_{s'})^2+(D_t-D_{t'})^2+(D_u-D_{u'})^2]
\nonumber
\eeq
The diagonal terms in $A$ are fixed simply as in all the other cases, while the non-diagonal terms depend on the type of configuration. They can be obtained by calculating the $k^2$ terms:
\beq
&&k_1^2+k_1^2+(k_2+k_3+k_4)^2+(k_2+k_3+k_4)^2+k_5^2+k_5^2\nonumber\\ &&~~=2(k_1^2+k_2^2+k_3^2+k_4^2+k_5^2)+4(k_2k_3+k_2k_4+k_3k_4)~.
\eeq
For $B$, the $k_1$ term is taken by the $pp':qq'$ state, the $k_2, k_3, k_4$ terms are taken by the $rr':ss'$ state and the $k_5$ term by the $tt':uu'$ state.  For $C$, it is simply written by using all the 12 states.  We can give the Gaussian integrals systematically, as shown here, but for the case in which the interactions and correlations are separable, the Gaussian integrals can be written as a product of various Gaussian integrals.  For an example of this case, we can write:
\beq
I^{(12:(34)^3:56)}(A,B,C)=I^{(12)}(A,B,C)\, I^{((34)^3)}(A,B,C)\, I^{(56)}(A,B,C)~.
\eeq

\nc\PTEP[1]{Prog.\ Theor.\ Exp.\ Phys.,\ \andvol{#1}} 

\end{document}